\documentclass[]{spie}  %>>> use for US letter paper
\usepackage{amsmath,amsfonts,amssymb}
\usepackage{graphicx}
\usepackage{subfig}
\usepackage[colorlinks=true, allcolors=blue]{hyperref}
\usepackage{siunitx}
\usepackage{float}
\DeclareSIUnit\angstrom{\text{Å}}
\DeclareSIUnit\photon{\text{ph}}
\DeclareSIUnit\arcsec{\text{arcsec}}
\usepackage{amssymb}
\usepackage{xcolor}
\usepackage{url}

\title{Progress on the simulation tools for the SOXS spectrograph: Exposure time calculator and End-to-End simulator}

\author[a]{M. Genoni}
\author[a]{A. Scaudo}
\author[b]{G. Li Causi}
\author[a]{L. Cabona}
\author[a,u]{M. Landoni}
\author[a]{S. Campana}
\author[c]{P. Schipani}
\author[d]{R. Claudi}
\author[a]{M. Aliverti}
\author[d]{A. Baruffolo}
\author[e]{S. Ben-Ami}
\author[f]{F. Biondi}
\author[c]{G. Capasso}
\author[g]{R. Cosentino}
\author[h]{F. D'Alessio}
\author[a]{P. D'Avanzo}
\author[e]{O. Hershko}
\author[i,k]{H. Kuncarayakti}
\author[l]{M. Munari}
\author[m,n]{G. Pignata}
\author[o]{A. Rubin} %SPIE Diverso
\author[l]{S. Scuderi} %SPIE Diverso
\author[h]{F. Vitali}
\author[o]{D. Young}
\author[p]{J. Achrén}
\author[q,n]{J. A. Araiza-Duran}
\author[r]{I. Arcavi}
\author[d]{F. Battaini}
\author[s]{A. Brucalassi} %SPIE Diverso -> Arceti
\author[e]{R. Bruch}
\author[d]{E. Cappellaro}
\author[c]{M. Colapietro}
\author[c]{M. Della Valle}
\author[d]{M. De Pascale}
\author[l]{R. Di Benedetto}
\author[c]{S. D'Orsi}
\author[e]{A. Gal-Yam}
\author[g]{M. Hernandez}
\author[i,k]{J. Kotilainen}
\author[c]{L. Marty}
\author[k]{S. Mattila}
\author[e]{M. Rappaport}
\author[d]{D. Ricci}
\author[a]{M. Riva}
\author[d]{B. Salasnich}
\author[o]{S. Smartt}
\author[l]{R. Zanmar Sanchez}
\author[t]{M. Stritzinger}
\author[g]{H. Ventura}

\affil[a]{INAF– Osservatorio Astronomico di Brera-Merate, via E. Bianchi 46, I-23807 Merate (LC), Italy;}
\affil[b]{INAF– Istituto di Astrofisica e Planetologia Spaziali, via Fosso del Cavaliere 100, Roma – Italy;}
\affil[c]{INAF – Osservatorio Astronomico di Capodimonte, Sal. Moiariello 16, I-80131, Naples, Italy;}
\affil[d]{INAF – Osservatorio Astronomico di Padova, Vicolo dell’Osservatorio 5, I-35122, Padua, Italy;}
\affil[e]{Weizmann Institute of Science, Herzl St 234, Rehovot, 7610001, Israel;}
\affil[f]{Max-Planck-Institut für Extraterrestrische Physik, Giessenbachstr. 1, D-85748 Garching, Germany;}
\affil[g]{FGG-INAF, TNG, Rambla J.A. Fernández Pérez 7, E-38712 Breña Baja (TF), Spain;}
\affil[h]{INAF – Osservatorio Astronomico di Roma, Via Frascati 33, I-00078 M. Porzio Catone, Italy;}
\affil[i]{Finnish Centre for Astronomy with ESO (FINCA), FI-20014 University of Turku, Finland;}
\affil[k]{Tuorla Observatory, Dept. of Physics and Astronomy, FI-20014 University of Turku, Finland;}
\affil[l]{INAF – Osservatorio Astroﬁsico di Catania, Via S. Soﬁa 78 30, I-95123 Catania, Italy;}
\affil[m]{Universidad Andres Bello, Avda. Republica 252, Santiago, Chile;}
\affil[n]{Millennium Institute of Astrophysics (MAS), Santiago, Chile;}
\affil[o]{Astrophysics Research Centre, Queen’s University Belfast, Belfast, BT7 1NN, UK;}
\affil[p]{Incident Angle Oy, Capsiankatu 4 A 29, FI-20320 Turku, Finland;}
\affil[q]{Centro de Investigaciones en Optica A. C., 37150 León, Mexico;}
\affil[r]{Tel Aviv University, Department of Astrophysics, 69978 Tel Aviv, Israel;}
\affil[s]{ESO, Karl Schwarzschild Strasse 2, D-85748, Garching bei München, Germany;}
\affil[t]{Aarhus University, Ny Munkegade 120, D-8000 Aarhus, Denmark;}
\affil[u]{INAF– Osservatorio Astronomico di Cagliari, via della Scienza 5, 09047, Selargius (CA), Italy;}

\authorinfo{send correspondence to: matteo.genoni@inaf.it}

\begin{document} 
\maketitle

\begin{abstract}
We present the progresses of the simulation tools, the Exposure Time Calculator (ETC) and End-to-End simulator (E2E), for the Son Of X-Shooter (SOXS) instrument at the ESO-NTT 3.58-meter telescope. The SOXS will be a single object spectroscopic facility, made by a two-arms high-efficiency spectrograph, able to cover the spectral range 350-2000 nanometer with a mean resolving power R$\approx$4500. While the purpose of the ETC is the estimate, to the best possible accuracy, of the Signal-to-Noise ratio (SNR), the E2E model allows us to simulate the propagation of photons, starting from the scientific target of interest, up to the detectors. We detail the ETC and E2E architectures, computational models and functionalities. The interface of the E2E with external simulation modules and with the pipeline are described, too. Synthetic spectral formats, related to different seeing and observing conditions, and calibration frames to be ingested by the pipeline are also presented.
\end{abstract}

\section{INTRODUCTION}
\label{sec:intro}  % \label{} allows reference to this section
Exposure Time Calculators (ETC) and End-to-End simulators (E2E) are valuable tools that can provide multiple functionalities at different stages of the instrument building. 
The purpose of ETCs is to provide the team and scientists with a fast, reliable, and easy-to-use tool that can produce valuable information about the instrument in relation to a specific astronomical observation, the signal-to-noise ratio (SNR) is an example.
End-to-End instrument models (E2E) are numerical simulators, which aim at simulating the expected astronomical observations starting from the radiation of the scientific sources (or calibration sources in the case of calibration frames) to the raw-frame data produced by instruments. Synthetic raw-frames can be ingested by the Data Reduction Software (DRS) to be analyzed in order to assess if top-level scientific requirements (such as spectral resolution, SNR) related to the specific science drivers, are satisfied with the specific instrument design. 
E2Es have been valuable software exploited in many types of astronomical instruments for different purposes. Specifically, they have been used in design phases to optimize and improve specific hardware components and parameters\cite{MOONS}, or for early verification of instrument performance\cite{Radial_vel_error_budget}. From the scientific point of view, they have been extensively exploited for assessing the feasibility of particularly challenging observations\cite{iLocater_e2e_drs}. Furthermore, instrument simulators are systematically exploited to aid both Data Reduction\cite{websim} and Data Analysis Software development, as well as for testing and verifying of existing data reduction pipelines. In fact, in the specific case of SOXS, the E2E simulator, which will be here presented, has a key goal to drive the optimization of the DRS development. In addition, it will also be a reliable tool to help the setting of calibration procedures and observation plans.
%SOXS will be a spectroscopic facility, made by two arms high efficiency spectrographs, able to cover the spectral range 350-2000 nanometer with resolving power R$\approx$4500. It is in construction phase and will be installed at the ESO-NTT 3.5-\unit{\meter} telescope (see for a wide overview\cite{soxs, soxs_update}).
%\textcolor{red}{

This paper is organized as follows, in Section \ref{sec:instrument}, an overview of the SOXS instrument is given. The Simulator, along with its architecture, simulated frames and iteration with Data Reduction Software is discussed in Section \ref{sec:e2e} and the Exposure Time Calculator is described in Section \ref{sec:etc}.
%}

\section{The SOXS Instrument}
\label{sec:instrument}
SOXS is a wide-band spectrograph for the ESO-NTT in La Silla (it will be installed at one of the Nasmyth foci of the NTT) covering in a single exposure the spectral range from the UV to the NIR (350-2000 nanometer). Its central structure (the common path) supports two distinct spectrographs, one operating in the UV-VIS 350-850 nanometer and the other in the NIR 800-2000 nanometer wavelength ranges, respectively. Both spectrographs can operate independently at different resolutions according to the slit widths: R$\approx$10000 with $0.5"$ slit, R$\approx$4500 with $1"$ slit and R$\approx$3300 with $1.5"$ slit. See general overview here \cite{soxs_update, soxs_gen_2022}.

%\begin{figure} [htp]
%\begin{center}
%\begin{tabular}{c} %% tabular useful for creating an array of images 
%\includegraphics[width=0.7\linewidth]{Images/The_SOXS_Instrument/soxs_instrument.png}
%\end{tabular}
%\end{center}
%\caption[example] 
%%>>>> use \label inside caption to get Fig. number with \ref{}
%{ \label{fig:soxs_cp} 
%SOXS Common-Path complete optical layout. On the left (in blue) the UV-VIS arm, on the right (in red) the NIR arm.}
%\end{figure}

The two arms are fed by the light coming through a common opto-mechanical system (the Common Path \cite{soxs_cp}). It redirects the light from the telescope focus to the spectrograph slits through relay optics reducing the F/number (from F/11 to F/6) and compensating for the atmospheric dispersion (only in the UV-VIS). The Common Path provides also the mechanism to drive the light to/from the other instrument subsystems, i.e. the acquisition camera and the calibration unit.

The Acquisition Camera subsystem has different functions: acquisition of the target for the spectrographs, monitoring of spectrographs co‐alignment and light imaging. It consists of a moving stage with positions enabling the various functions, a folding mirror, a filter wheel ({\it u, g, r, i, z, y} LSST and V Johnson bands) front of a compact camera which relays the Nasmyth focus (field of view of $3.5\times 3.5$ arcmin) on the  13.0 $\mu$m, $1024\times1024$ pixel detector CCD. See for details \cite{acq_guid_soxs}.

\begin{figure} [ht]
\begin{center}
\begin{tabular}{c} %% tabular useful for creating an array of images 
\includegraphics[width=0.35\linewidth]{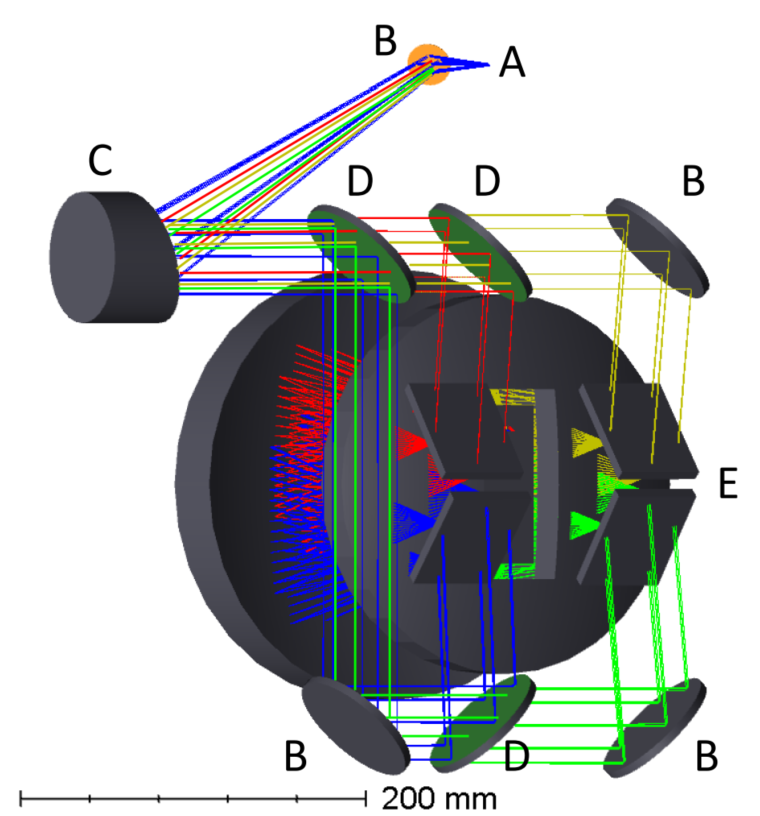}
\end{tabular}
\end{center}
\caption[example] 
%>>>> use \label inside caption to get Fig. number with \ref{}
{ \label{fig:soxs_vis} 
SOXS UV-VIS spectrograph complete optical layout. A. slit plane, B. reflective mirrors, C. OAP Collimator, D. dichroic filters, E. gratings.}
\end{figure}

\begin{figure} [ht]
\begin{center}
\begin{tabular}{c} %% tabular useful for creating an array of images 
\includegraphics[width=0.55\linewidth]{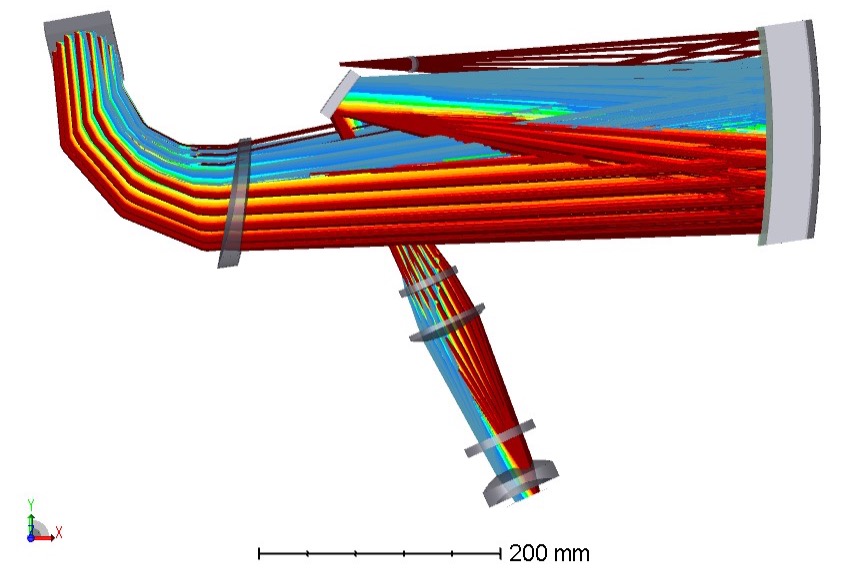}
\end{tabular}
\end{center}
\caption[example] 
%>>>> use \label inside caption to get Fig. number with \ref{}
{ \label{fig:soxs_nir} 
SOXS NIR spectrograph complete optical layout. The optical beams goes the double pass collimating and cross-dispersing optics. The main disperser is a $44\deg$ blaze angle echelle grating. A field mirror redirect the beam down to the camera.}
\end{figure}

The UV-VIS spectrograph arm is based on a novel multi-grating concept (for details\cite{MITS} and Fig. \ref{fig:soxs_vis}), in which the incoming beam is partitioned into four polychromatic beams using dichroic surfaces, each covering a waveband range of $\sim 100$ nanometer, named as quasi-order. Each quasi-order is diffracted by an ion-etched grating. The detector, located 4 mm behind the camera field flattener back surface, is an e2V CCD44-82 CCD ($2048\times 4096$ pixels, pixel size 15 micrometer, see for details\cite{soxs_vis}).
% \newline
The near infrared spectrograph, showed in Fig. \ref{fig:soxs_nir}, is a cross-dispersed echelle, with R$\sim 5000$ (for $1''$ slit), covering the wavelength range from 800 to 2000 nanometer with 15 orders (see for details\cite{soxs_optical_design}). It is based on the 4C concept, characterized by a very compact layout, reduced weight of optics and mechanics, and good stiffness. The spectrograph is composed of a double-pass collimator and a refractive camera, an R-1 grating as main disperser and a prism-based cross disperser. The detector is a Teledyne H2RG array operated at 40K, $2048\times 2048$ pixels, pixel size 18 micrometer (see for details\cite{soxs_nir_paper}).
\newline


The calibration unit, provides the calibration spectra to remove the instrument signature. The calibration spectra are generated using a synthetic light source, adopting an integrating sphere equipped with lamps suitable for wavelength and flux calibrations across the full wavelength range of the instrument (350-2000 nanometer); for details see\cite{soxs_cal}. The following lamps are used:
\begin{itemize}
  \item Quartz-tungsten-halogen (QTH) lamp, for flat field frames 500-2000 nanometer;
  \item Deuterium (D2) lamp, for flat field 350-500 nanometer (used simultaneously with QTH lamp for UV-VIS arm);
  \item ThAr hollow cathode lamp, for UV-VIS wavelength calibration;
  \item Ne-Ar-Hg-Xe pen-ray lamps bundled together, for NIR wavelength calibration. The individual lamps are controlled to operate together as one lamp.
\end{itemize}

The instrument is now under intense AIT phase in the consortium premises (Astronomical Observatories of Padova and Merate, Italy) as detailed in several contributions \cite{soxs_gen_2022,soxs_cp_2022,soxs_acq_2022,soxs_nir_2022}.

%\newpage
\section{End-to-End simulator}
\label{sec:e2e}
\begin{figure} [htp]
\begin{center}
\begin{tabular}{c} %% tabular useful for creating an array of images 
\includegraphics[width=0.75\linewidth]{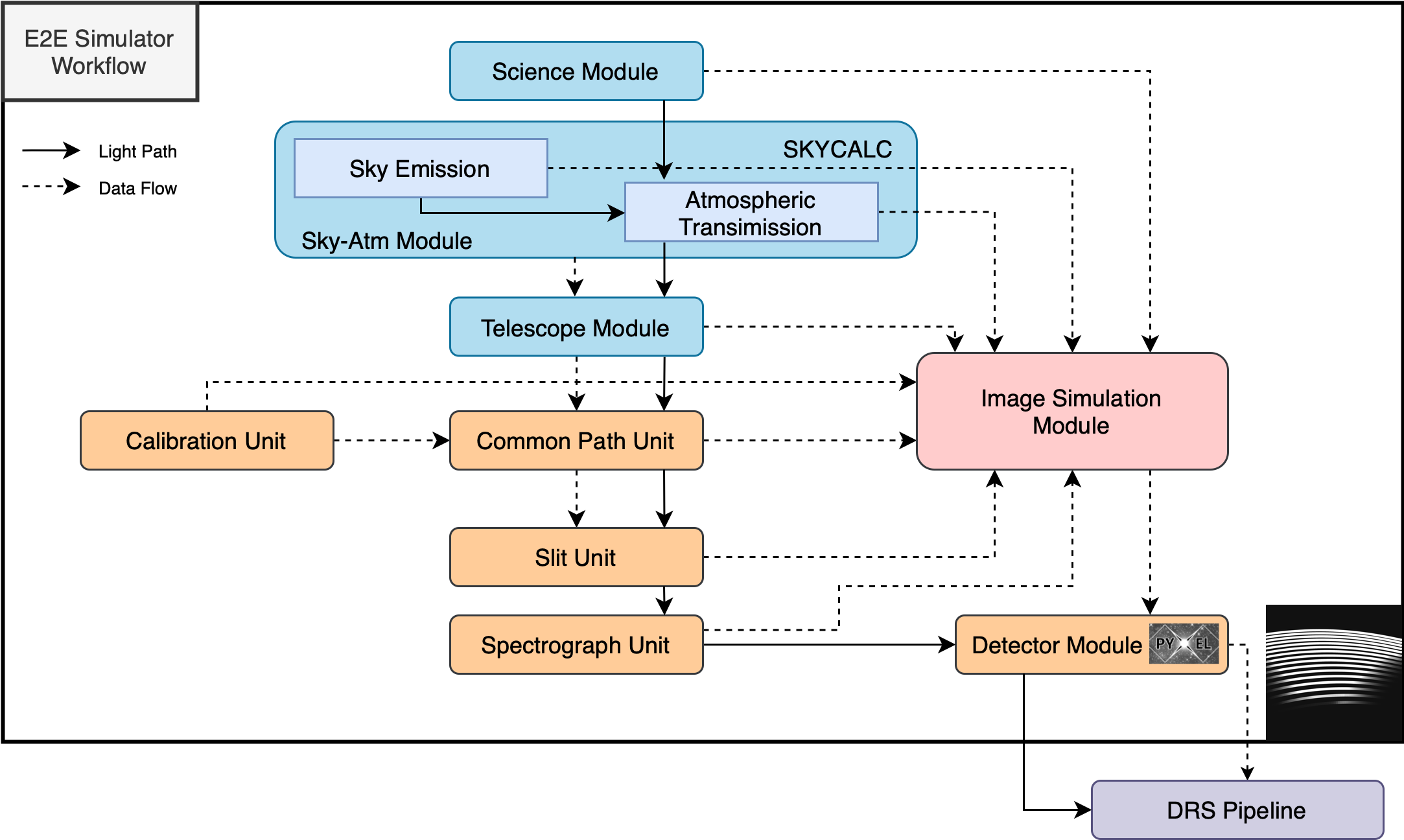}
\end{tabular}
\end{center}
\caption[example] 
%>>>> use \label inside caption to get Fig. number with \ref{}
{ \label{fig:e2e_arc} 
E2E simulator workflow schematic description. The solid arrows represent the real light path, while the dashed arrows show: the simulation data-flow, how the different modules and units are interfaced and their connection to the simulator core, which is the Image Simulation Module. Orange blocks are units of the Instrument Module, while blue blocks are related to simulation modules independent from the specific instrument.}
\end{figure}

The End-to-End simulator architecture is highly modular, composed by different modules (each one with specific tasks and functionalities) units and interfaces, as described in the schematic workflow of Fig. \ref{fig:e2e_arc}. 
Modularity and flexibility are key points in the definition of tasks and interfaces among different modules or units to allow this tool for being scalable and adaptable to different kind of simulations (science versus calibrations frames) and spectroscopic instrumentation. 
%Modules and Units are characterized by the main tasks for which they are in charge, the required inputs and expected outputs (in a specified format). 
The simulator is written in Python 3.9 and uses specific libraries for specific functionalities and for interfacing with other software, like commercial optical ray tracing Zemax-OpticStudio®.

\subsection{The Simulator Architecture}
\label{sec:e2e_architecture}
In the following a brief description of the modules and units showed in the workflow diagram is given; more details can be found here \cite{soxs_e2e_2020}. 

\subsubsection{Science Module}
\label{sec:e2e_architecture_sci_module}
\begin{figure} [ht]
\begin{center}
\begin{tabular}{c} %% tabular useful for creating an array of images 
\includegraphics[width=0.55\linewidth]{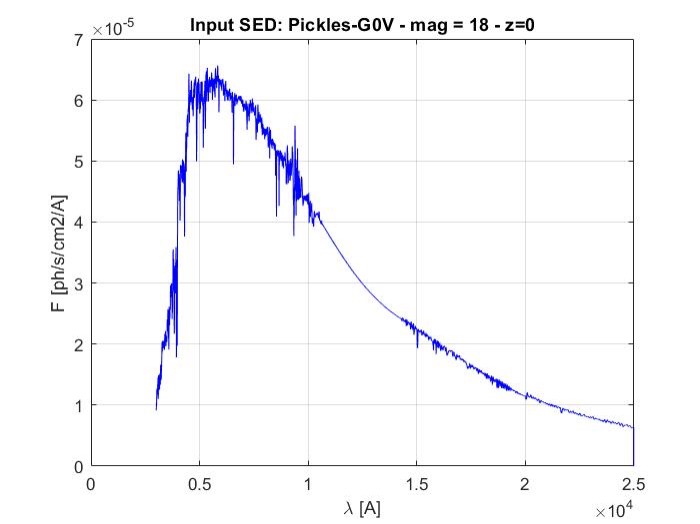}
\end{tabular}
\end{center}
\caption[example] 
%>>>> use \label inside caption to get Fig. number with \ref{}
{ \label{fig:soxs_sci_module} 
Example of a synthetic 1D spectrum for a G0V star (Pickels model) sampled at resolution of 0.5 \AA, mag 18 (V band of Johnson-Cousins filters, in Vega magnitude system) and at $z=0$.}
\end{figure}

The Science Object Module has the purpose to generate a synthetic 1D spectrum related to a specific astronomical source, at a resolution higher than the one selected for the instrument simulation (which is related to the instrumental resolution dependent on the slit width). 
The 1D spectral flux density distribution is obtained by a set of parameters (e.g. object type - spectral type for stars -, magnitude and redshift) or by loading the spectrum from a user-defined library. 
%Once the spectrum is loaded, it is first shifted in wavelength according to the redshift, then it is normalized at the reference wavelength of the provided magnitude pass-band and then re-scaled accordingly. 
%The output of this module is a FITS table containing the spectral flux density, in units of photon flux, at each wavelength. 
An example of a synthetic 1D spectrum for a G0V star (Pickels model) sampled at resolution of 0.5  Angstrom , is shown in the Fig. \ref{fig:soxs_sci_module}.

\subsubsection{Sky-Atmosphere Module}
\label{sec:e2e_architecture_sky_module}
The task of this module is to model the scattering, absorption and emission occurring in the Earth's atmosphere. 
This is done by calling sky-emission and atmospheric transmission spectra of a dedicated library built using the ESO SkyCalc tool (available at the web page\cite{skycalc}). %, which is based on the Cerro Paranal Advanced Sky Model. 
%These spectra can be loaded from the library according to the sky conditions parameters set for the simulation, i.e. moon phase, air-mass (AM) and precipitable water vapor (PWV). 
The sky radiance spectrum loaded is in units of photon second meter$^2$ micrometer arcsec$^2$, thus it is first calculated for the on-sky area related to the slit length and selected slit width according to the simulation (both in  arcsec ) and then the units are converted in{ photon second cm$^2$  Angstrom ).

\begin{figure} [ht]
\begin{center}
\begin{tabular}{c} %% tabular useful for creating an array of images 
\includegraphics[width=0.7\linewidth]{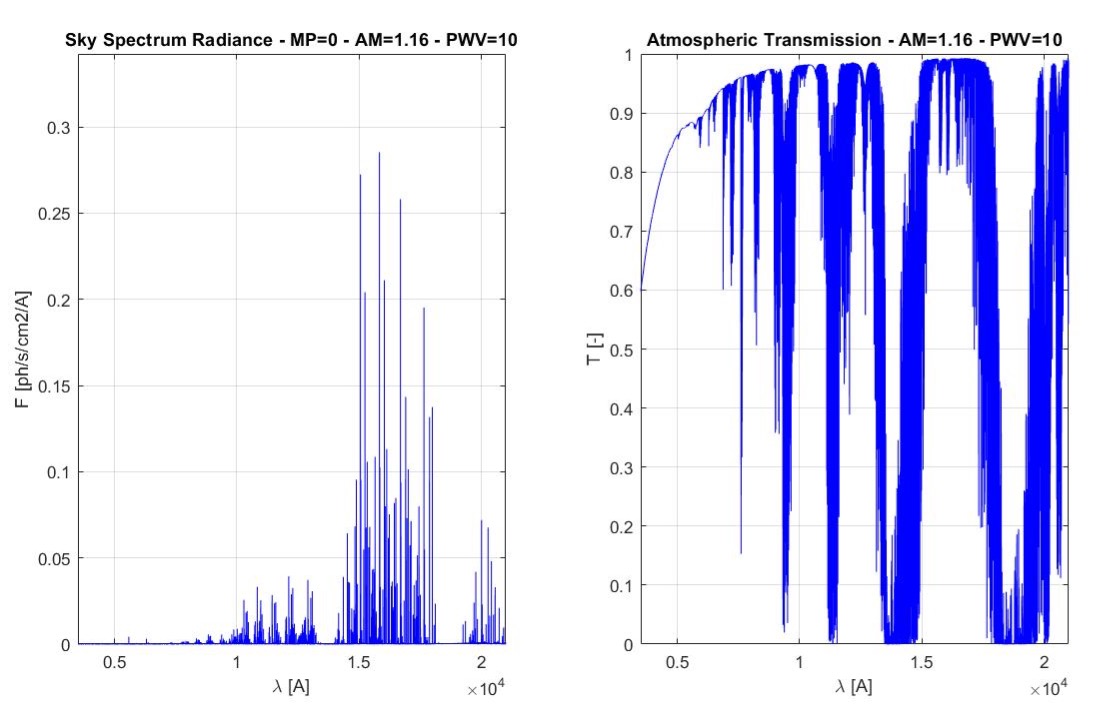}
\end{tabular}
\end{center}
\caption[example] 
%>>>> use \label inside caption to get Fig. number with \ref{}
{ \label{fig:soxs_sky_module} 
Example of a sky radiance spectrum and atmospheric transmission calculated through the ESO-SkyCalc tool. New moon, air-mass=1.16 and precipitable water vapor (PWV) = 10  mm }
\end{figure}

%\newpage
\subsubsection{Telescope Module}
The aim of this module is to predict the telescope throughput based on the available telescope mirrors reflectivity data. For the NTT a global telescope throughput of 0.61 is considered, given the three reflections along the telescope optical path.
\label{sec:e2e_architecture_tel_module}

%\newpage
\subsubsection{Calibration Unit}
\label{sec:e2e_architecture_cal_unit}
The calibration unit is in charge of simulating the Spectral Energy Distribution (SED) of the calibration sources. This unit requires to know which kind of lamps are to be simulated flats lamp or ThAr and pen-rays lamps, according to the specific calibration frame to be generated, the instrument resolution and the type of calibration mask. In fact, SOXS foresees a single pinhole and a multi pinholes (9 pinholes) mask for calibrations and DRS purposes. Examples of the QTH spectrum (in UV-VIS) and penrays (Ar-Hg-Ne-Xe) spectrum in the NIR are shown in Fig. \ref{fig:soxs_cal_unit}.

\begin{figure} [h!]
\begin{center}
\begin{tabular}{c} %% tabular useful for creating an array of images 
\includegraphics[width=0.45\linewidth]{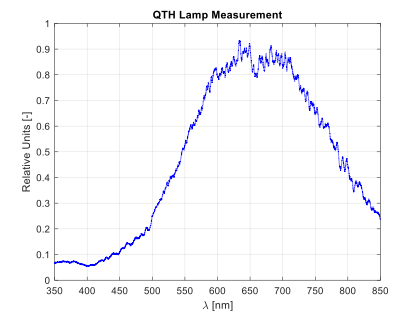}
\includegraphics[width=0.45\linewidth]{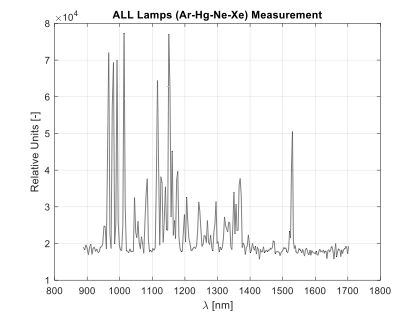}
\end{tabular}
\end{center}
\caption[example] 
%>>>> use \label inside caption to get Fig. number with \ref{}
{ \label{fig:soxs_cal_unit} 
Example of calibration sources spectra used in SOXS. Left panel: QTH spectrum, right panel: penrays. Both spectra have been measured by the Calibration Unit work-package team.}
\end{figure} 

\subsubsection{Common Path Unit and Slit Unit}
\label{sec:e2e_architecture_cp_slit_unit}
The task of the Common-Path (CP) unit is to predict the light distribution at the CP focal plane and throughput for both the UV-VIS arm and NIR arm. 
The required inputs are seeing value for the observing conditions, the CP image scale and the optics data regarding glasses, coatings and mirrors reflectivity, in order to estimate the image on the CP focal plane and the efficiency. 
%The UV-VIS and NIR common path arms efficiency has been calculated combining both measurements data from manufacturer test reports, for the already built optical elements, and estimates for the components that are currently under procurement.

\begin{figure} [h!]
\begin{center}
\begin{tabular}{cc} %% tabular useful for creating an array of images 
\includegraphics[width=0.49\linewidth]{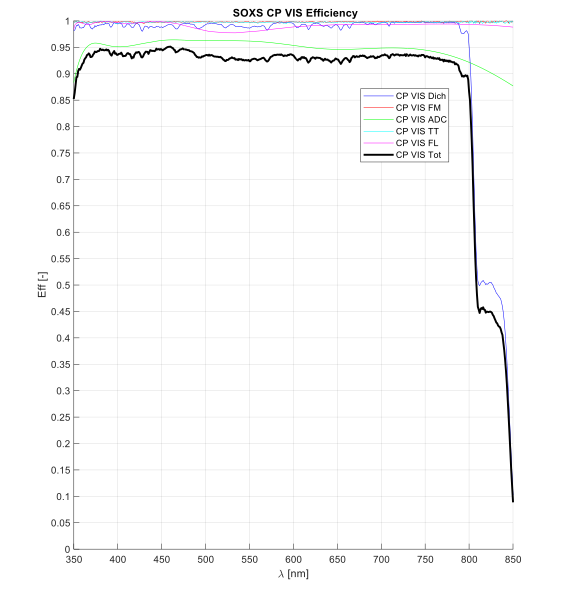}&
\includegraphics[width=0.49\linewidth]{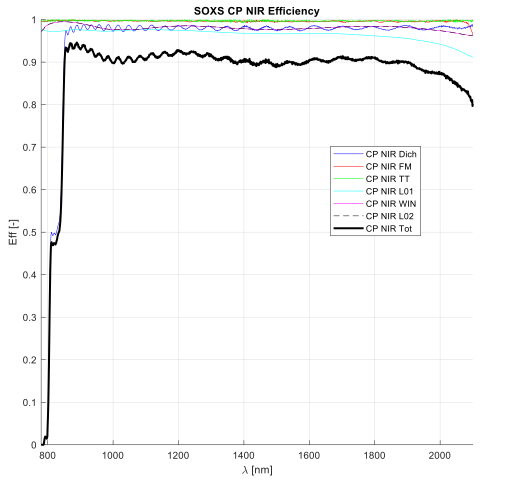}\\
\end{tabular}
\end{center}
\caption[example] 
%>>>> use \label inside caption to get Fig. number with \ref{}
{ \label{fig:soxs_cp_vis_unit} 
Left panel: common-Path UV-VIS arm efficiency, in black bold line, while in colored lines the single Common-Path optical elements efficiency. Dich – Dichroic, FM – folding mirror, ADC – atmospheric dispersion corrector, TT – tip-tilt mirror, FL – field lens. Right panel: common-Path NIR arm efficiency, in black bold line, while in colored lines the single Common-Path optical elements efficiency. Dich – Dichroic, FM – folding mirror, TT – tip-tilt mirror, L01 – lens 1 (refocuser), WIN – window, L02 – lens2 (field lens).}
\end{figure}

%\begin{figure} [ht]
%\begin{center}
%\begin{tabular}{c} %% tabular useful for creating an array of images 
%\includegraphics[width=0.4\linewidth]{Images/E2E/Architecture/common_path_eff_NIR.png}
%\end{tabular}
%\end{center}
%\caption[example] 
%>>>> use \label inside caption to get Fig. number with \ref{}
%{ \label{fig:soxs_cp_nir_unit} 
%Common-Path NIR arm efficiency, in black bold line, while in colored lines the single Common-Path optical elements efficiency. Dich – Dichroic, FM – folding mirror, TT – tip-tilt mirror, L01 – lens 1 (refocuser), WIN – window, L02 – lens2 (field lens).}
%\end{figure}

The slit-unit calculates the fraction of light passing through the UV-VIS and NIR slits, which are located on their respective common-path arms focal plane, from the variation of seeing in wavelength (w.r.t. the reference value set at 5000  Angstrom ), airmass, atmospheric turbulence parameters and considering the specific sizes (the slit length is constant $12"$, while the width changes according to the simulated set of instrument parameters).

%\begin{figure} [ht]
%\begin{center}
%\begin{tabular}{c} %% tabular useful for creating an array of images 
%\includegraphics[width=0.5\linewidth]{Images/E2E/Architecture/slit_eff.jpg}
%\end{tabular}
%\end{center}
%\caption[example] 
%%>>>> use \label inside caption to get Fig. number with \ref{}
%{ \label{fig:soxs_slit_eff} 
%Slit throughput profile for a seeing value of $1"$ (at reference wavelength of 5500  Angstrom ) and the $1"$ slit width (for which the SOXS resolving power is about 4500).}
%\end{figure}

\subsubsection{Spectrograph Unit}
\label{sec:e2e_architecture_spec_unit}
The purpose of this unit is to simulate the physical effects of the different optical components of the spectrometers with the final aim of predict the echellogram (spectral format) at the UV-VIS and NIR focal plane, the throughput and a database of Point Spread Function (PSF) maps for a set spectral resolution element (which are used by the Image Simulation Module for rendering the synthetic frames). 
%Aberrations, distortion and diffraction effects have already been taken into account at this current simulator version, while the physical operative conditions of the instrument in term of mechanical and thermal effects will be introduced in the future development versions according to the required simulation scenarios.
\newline
%As for the common path, also the spectrograph arms efficiency is calculated combining both measurements data from manufacturer test reports, for the already built optical elements, and estimations for the components that are currently under procurement. UV-VIS is plotted in Fig. \ref{fig:soxs_eff}, while NIR in Fig. \ref{fig:soxs_eff}.
The UV-VIS quasi-orders and NIR orders efficiency (from spectrograph entrance to focal plane, detector QE excluded) are shown in Fig. \ref{fig:soxs_eff}.

\begin{figure} [ht]
\begin{center}
\begin{tabular}{cc} %% tabular useful for creating an array of images 
\includegraphics[width=0.5\linewidth]{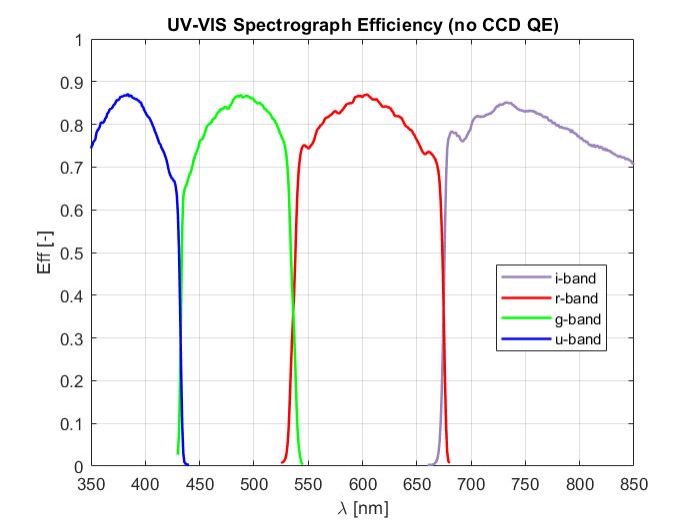}&
\includegraphics[width=0.5\linewidth]{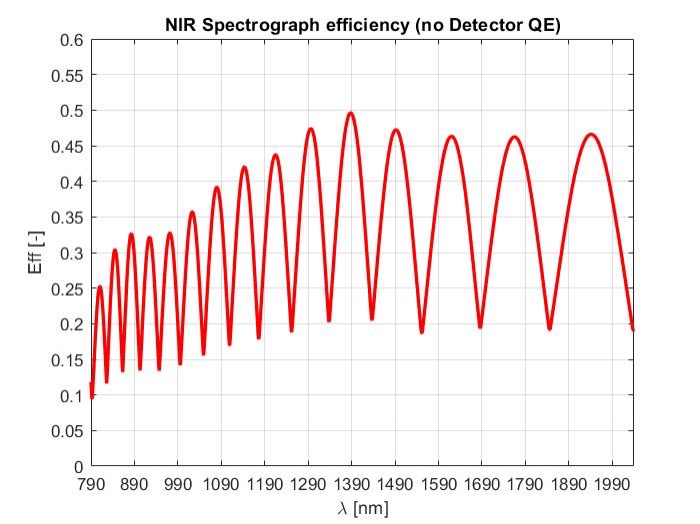}\\
\end{tabular}
\end{center}
\caption[example] 
%>>>> use \label inside caption to get Fig. number with \ref{}
{ \label{fig:soxs_eff} 
Spectrograph efficiency profiles (detector QE excluded) Left panel: UV-VIS spectrograph quasi orders. Right panel: NIR.}
\end{figure}

%\begin{figure} [ht]
%\begin{center}
%\begin{tabular}{c} %% tabular useful for creating an array of images 
%\includegraphics[width=0.5\linewidth]{Images/E2E/Architecture/nir_eff.jpg}
%\end{tabular}
%\end{center}
%\caption[example] 
%>>>> use \label inside caption to get Fig. number with \ref{}
%{ \label{fig:soxs_nir_eff} 
%NIR spectrograph global efficiency curve.}
%\end{figure}

The PSF maps database is computed by directly querying the optical design ray-tracing files (built with commercial optical design software Zemax-OpticStudio®), using specific API-based software codes to extract the required data for a defined set of wavelengths along each diffraction order (or the for quasi-orders in the UV-VIS arm). 
In particular, the PSF maps are calculated using the Huygens PSF analysis tool of Zemax-OpticStudio®, which includes diffraction and optical aberrations. 
Being SOXS a long slit spectrograph, the PSF map for each database wavelength can be computed both from an average of the PSF maps, 64micrometer wide, related to five positions along the slit (-0.662, -0.331, 0, 0.331, 0.662  mm , corresponding to -6, -3, 0, 3, 6  arcsec ), and by interpolating along slit direction the map of these five positions, according to the required simulation. 
Examples for both arms, first UV-VIS then NIR, in  Fig. \ref{fig:soxs_vis_psf}.

\begin{figure}[htp]
\begin{center}
\captionsetup[subfloat]{position=top,labelformat=empty}
\subfloat[]{\includegraphics[width=0.5\linewidth]{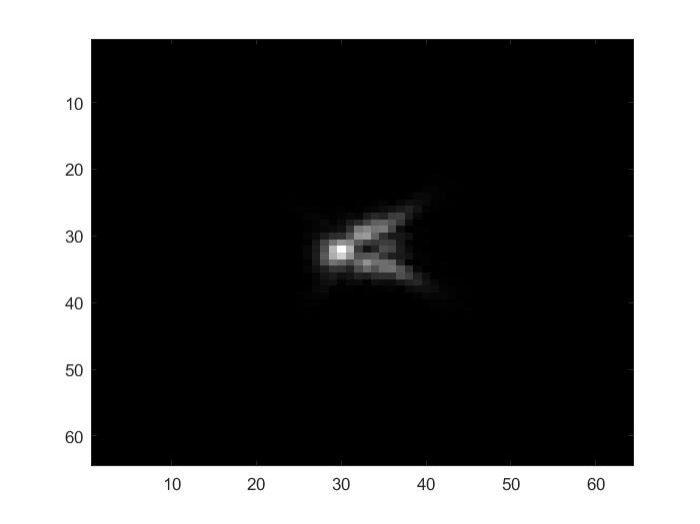}}
\subfloat[]{\includegraphics[width=0.5\linewidth]{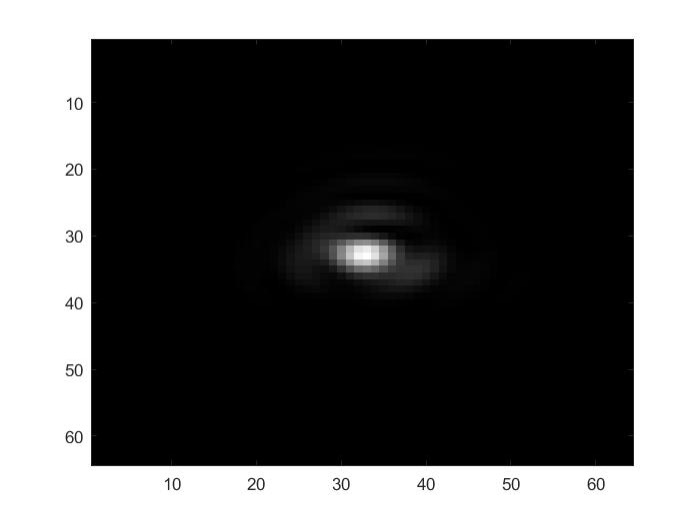}}

%\subfloat[]{\includegraphics[width=0.5\linewidth]{Images/E2E/Architecture/psf/psf_vis_center_left.jpg}}
%\subfloat[]{\includegraphics[width=0.5\linewidth]{Images/E2E/Architecture/psf/psf_vis_center_right.jpg}}

%\subfloat[]{\includegraphics[width=0.5\linewidth]{Images/E2E/Architecture/psf/psf_vis_down_left.jpg}}
%\subfloat[]{\includegraphics[width=0.5\linewidth]{Images/E2E/Architecture/psf/psf_vis_down_right.jpg}}
\caption[example] 
%%>>>> use \label inside caption to get Fig. number with \ref{}
{ \label{fig:soxs_vis_psf} 
Example of PSF maps, 64 micrometer wide, computed for center slit position. Left panel: UV-VIS spectrograph. Right Panel: NIR spectrograph.}
\end{center}
\end{figure}

The spectral format data, required to produce the synthetic echellograms by the Image Simulation Module, are the diffraction order number (or quasi order id, for UV-VIS), wavelength (the same set used in the computation of the PSF maps database), centroid coordinates of both the central slit position and edges (through which the slit image tilt on the focal plane is computed) and slit image size (i.e. the geometrical full width in main dispersion and spatial direction).
%\newline
%\newline

The full database, comprised of efficiency, PSF maps and spectral format, produced by the Spectrograph Unit is stored for the usage of the Image Simulation Module.

\subsubsection{Image Simulation Module}
\label{sec:e2e_architecture_img_module}
This portion of the simulator is the kernel of the whole system and put together the outputs of all the other modules and units. This piece of software is responsible for rendering the photons distribution of each spectral resolution element for each order as should be detected at the level of spectrographs focal plane by the detectors. 
In particular, the module first interpolates on a sub-pixel scale, of which the oversampling can be set according to the required simulation accuracy, all the instrumental data regarding wavelength, image centroid coordinates, sampling (in both main dispersion and spatial directions), slit image tilt, average PSF map and efficiency. 
Then, it produces the photons distribution of each wavelength in sub-pixel scale by convolving the slit (long slit or pinhole masks according to the specific science or calibration frame type) image with the corresponding PSF map. The spectral slit images are properly sized and tilted according to the sampling and tilt variation along each order, and scaled for the integrated spectral flux and efficiency. 
The code architecture has been developed in order to properly exploit the functionality of {\tt numba}\cite{numba} and to take advantage of the parallel features of the concurrent Python package, such that the different diffraction orders (or a fraction of them) can run in parallel. 
Once all the spectral images have been piled up to sum their photons distribution at sub-pixel scale, the synthetic frame is re-binned to pixel scale. Then, the photon noise and all other specific detector noises are added by the Detector Module.

\subsubsection{Detector Module}
%\textcolor{red}{
%
The detector module exploits the Pyxel \cite{pyxel_2020}
%\textcolor{red}{[REFERENCE]} 
simulation framework to model the different detector effects on the rendered data at pixel scale. This tool works through a series of modules that are executed in cascade simulating the whole detection chain. 
For example, at the photon generation level, it is possible to model the shot noise, while the quantum efficiency (QE), dark current (DC), and the effect of cosmic ray are added at the charge generation level. Several other effects can be added at different levels of the simulation, including read-out noise (RON), bias levels, fix pattern noise caused by pixels non-uniform response (PRNU), pixel cross-talk (due to charge diffusion) and more. 
In addition, for NIR detectors a specific module, the HxRG noise generator, allows to model the kTC bias noise, white readout noise and alternating column noise (ACN). Through this framework, it is also possible to generate bias and dark frames for calibrations.
%}
The QE curves for both the UV-VIS CCD and the NIR detector implemented in the simulation are shown in Fig. \ref{fig:qe}.

\begin{figure} [h!]
\begin{center}
\begin{tabular}{c} %% tabular useful for creating an array of images 
\includegraphics[width=0.5\linewidth]{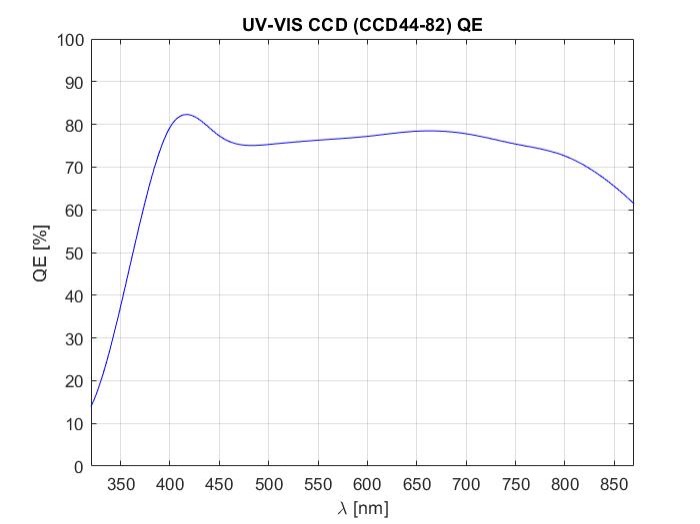}
\includegraphics[width=0.5\linewidth]{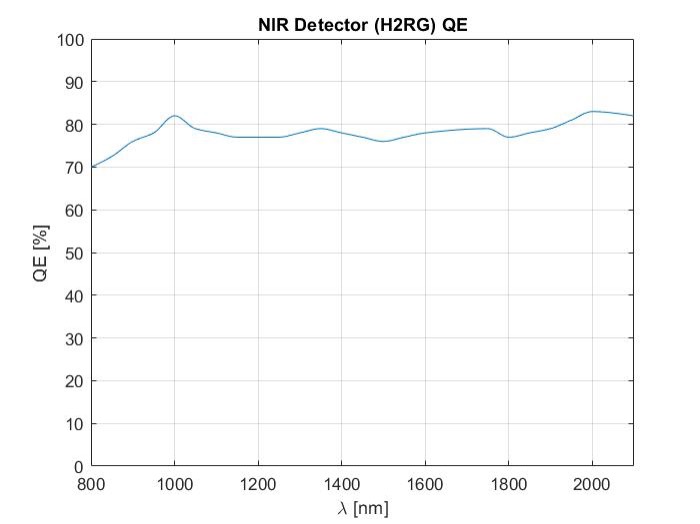}
\end{tabular}
\end{center}
\caption[example] 
%>>>> use \label inside caption to get Fig. number with \ref{}
{ \label{fig:qe} 
Spectrograph detectors QE. Left panel: UV-VIS CCD. Right panel NIR H2RG detector.}
\end{figure} 

The parameters for some relevant implemented effects (extrapolated from datasheet, measurements by manufacturer tests or from same detector type in similar instrument) are the followings:

\begin{table}[ht]
\label{tab:detector}
\begin{center}       
\begin{tabular}{|l|l|l|l|l|l|}
\hline
\rule[-1ex]{0pt}{5ex}  & Gain (ADU/e-) & RON (rms e-) & DC (e/pix/s) & PNUR (\%) & Cross-talk (\%)  \\
\hline
\rule[-1ex]{0pt}{5ex} UV-VIS CCD & 1.25 & 2.3 & $2.7\times 10^{-5}$ & 1.6 & 1   \\
\hline
\rule[-1ex]{0pt}{5ex} NIR Detector & 0.5 & 11 (CDS reading) & $<10^{-3}$ & 4 & 0.5   \\
\hline
\end{tabular}
\end{center}
\end{table}
\label{sec:e2e_architecture_dec_module}

\subsection{Simulated Frames}
\label{sec:e2e_frames}
In this section we present some examples of simulated frames generated with the E2E simulator. A first example (see Fig. \ref{fig:m2v}) is the NIR synthetic image of a M2V (Pickles) stellar spectral type with \textit{V}-mag = 18 (Vega System) observed at new moon with air-mass = 1, precipitable water vapor (PWV) = 10  mm , seeing = 1  arcsec  (at 5000  Angstrom ) and with the selected slit width of 1  arcsec . The total exposure time is 900 seconds realized through a single exposure. A simple set of detector noises was added via Pyxel, such as shot noise, dark current (DC) and read out noise (RON). The different diffraction orders traces with the object centered in the slit and sky emission features are clearly visible.

\begin{figure} [h!]
\begin{center}
\begin{tabular}{c} %% tabular useful for creating an array of images 
\includegraphics[width=0.5\linewidth]{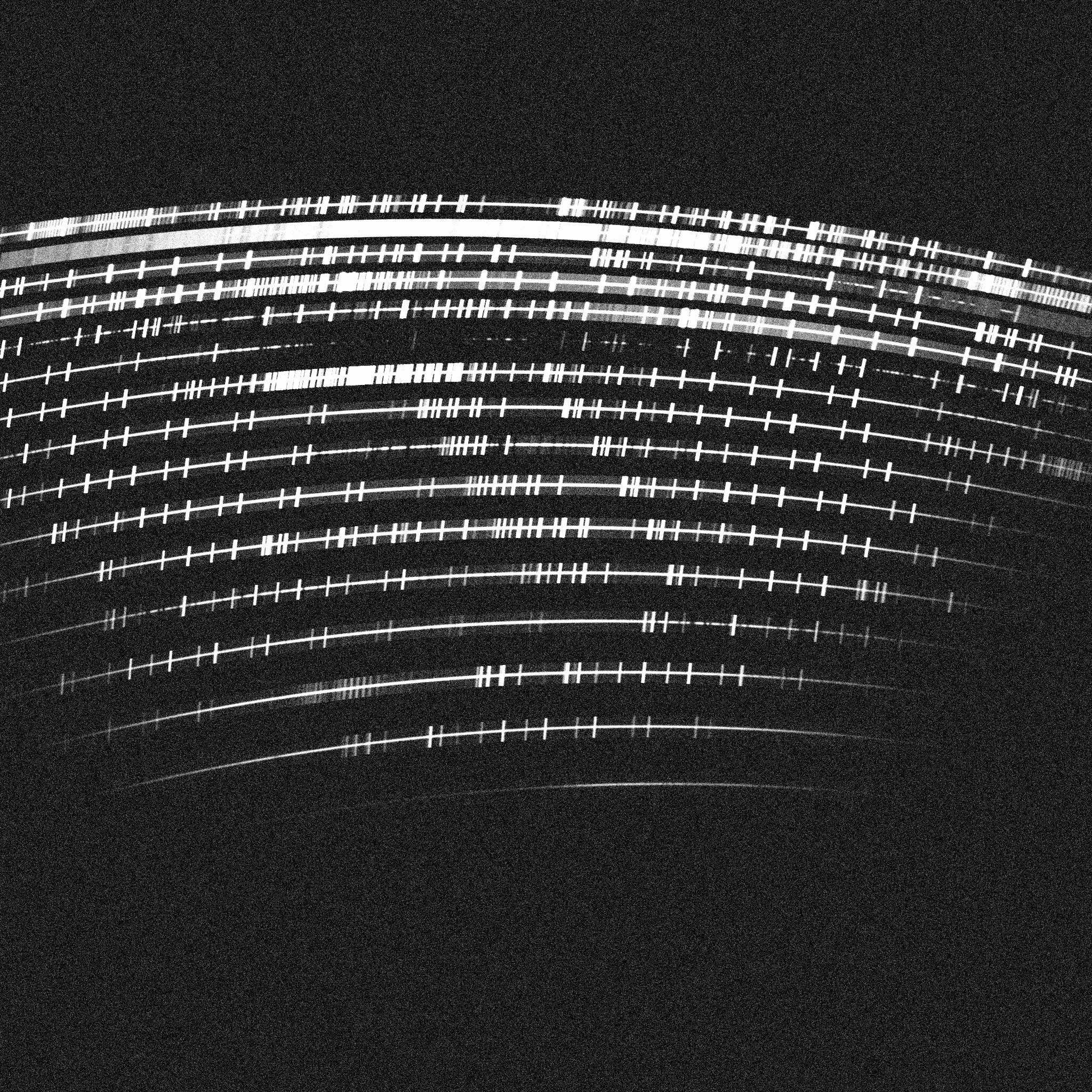}
\end{tabular}
\end{center}
\caption[example] 
%>>>> use \label inside caption to get Fig. number with \ref{}
{ \label{fig:m2v} 
NIR arm synthetic science frame of a M2V (Pickles) stellar spectral type with \textit{V}-mag = 18 (Vega System).}
\end{figure}

%\newpage
The images in Fig. \ref{fig:e2e_calib} and Fig. \ref{fig:e2e_calib_mph} represent three different NIR wavelength calibration frames taken with 10 seconds of exposure that were used during the iteration with the Data Reduction Software, as discussed in the next section. 
In the first set of frames a flat field through the 1  arcsec  slit width and a second flat field through a single pinhole are simulated. 
The wavelength calibration frame of Fig. \ref{fig:e2e_calib_mph} simulates the emission lines of the PenRays (Ar-Hg-Ne-Xe) lamps through a multi-pinhole mask. The spectral resolution elements related to the emission lines of the lamps along each order can be seen both in the full frame and in the zoomed box, where also the 9 pinholes images in cross dispersion direction can be distinguished.

\begin{figure}[htp]
\begin{center}
\subfloat[Flat field through slit ]{\label{fig:a_lamp_slit}\includegraphics[width=0.425\linewidth]{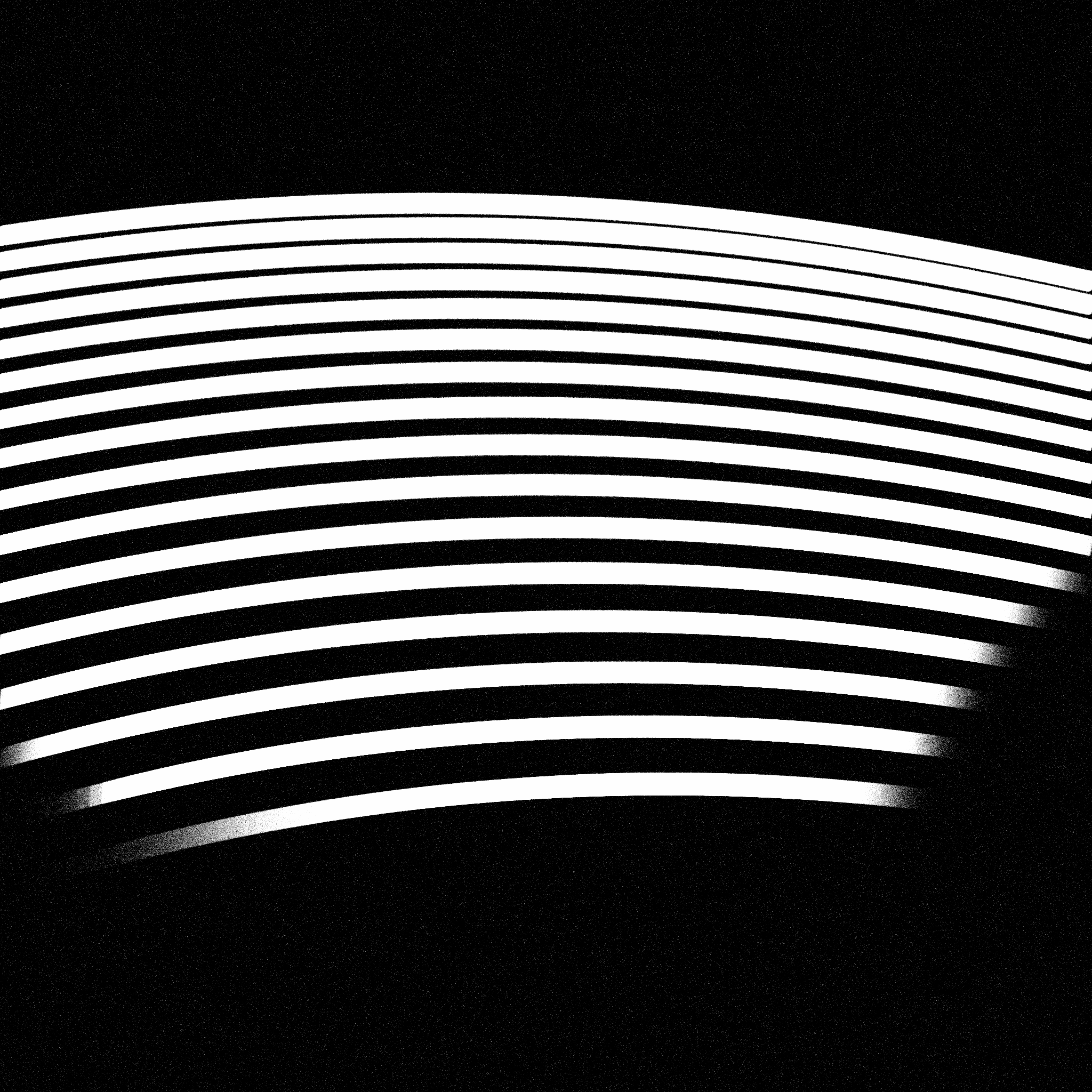}} \qquad
\subfloat[Flat field through single pinhole ]{\label{fig:b_flat_sph}\includegraphics[width=0.425\linewidth]{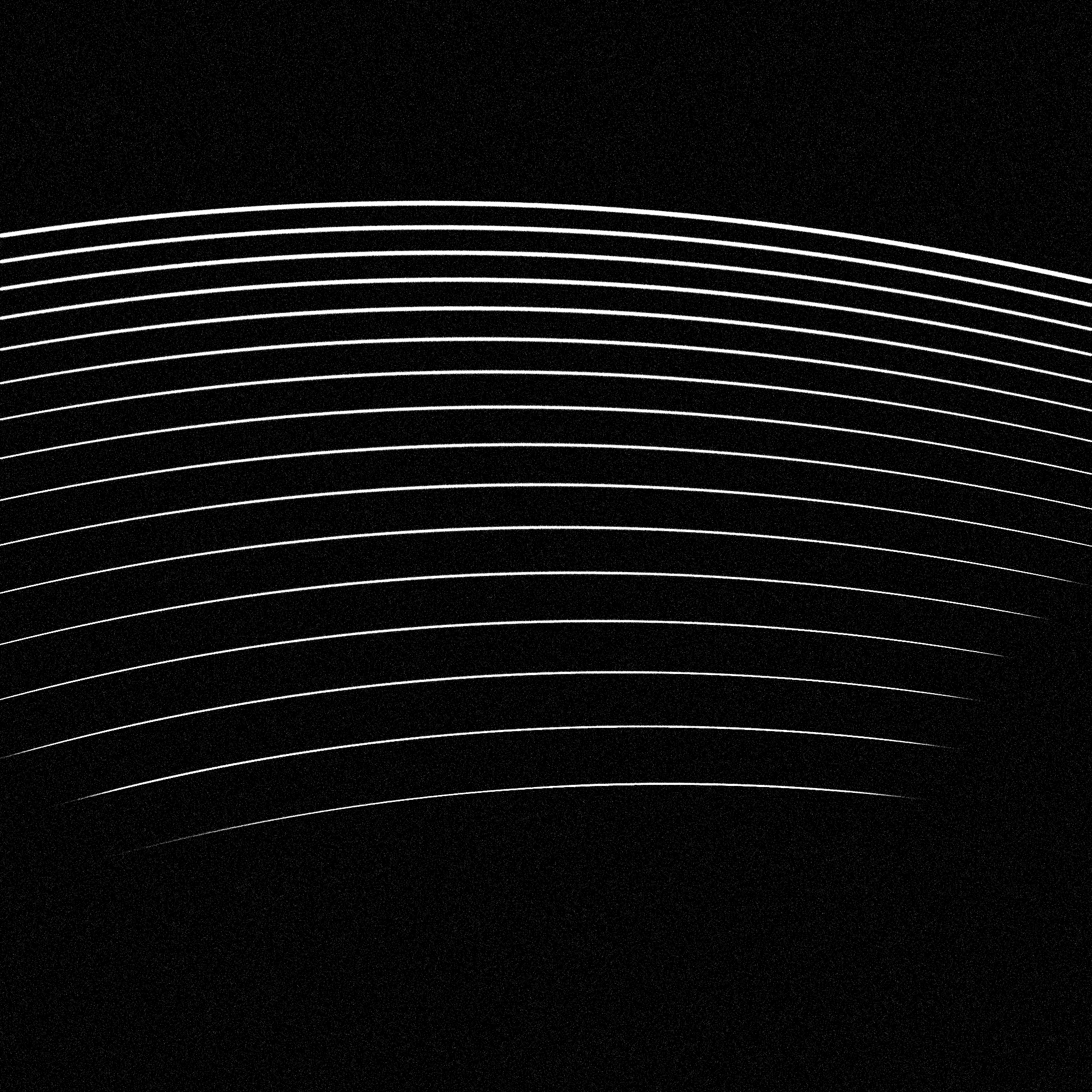}}

\caption[example] 
%>>>> use \label inside caption to get Fig. number with \ref{}
{ \label{fig:e2e_calib}  
NIR arm calibration frames.
}
\end{center}
\end{figure}

\begin{figure} [h!]
\begin{center}
\begin{tabular}{c} %% tabular useful for creating an array of images 
\includegraphics[width=0.9\linewidth]{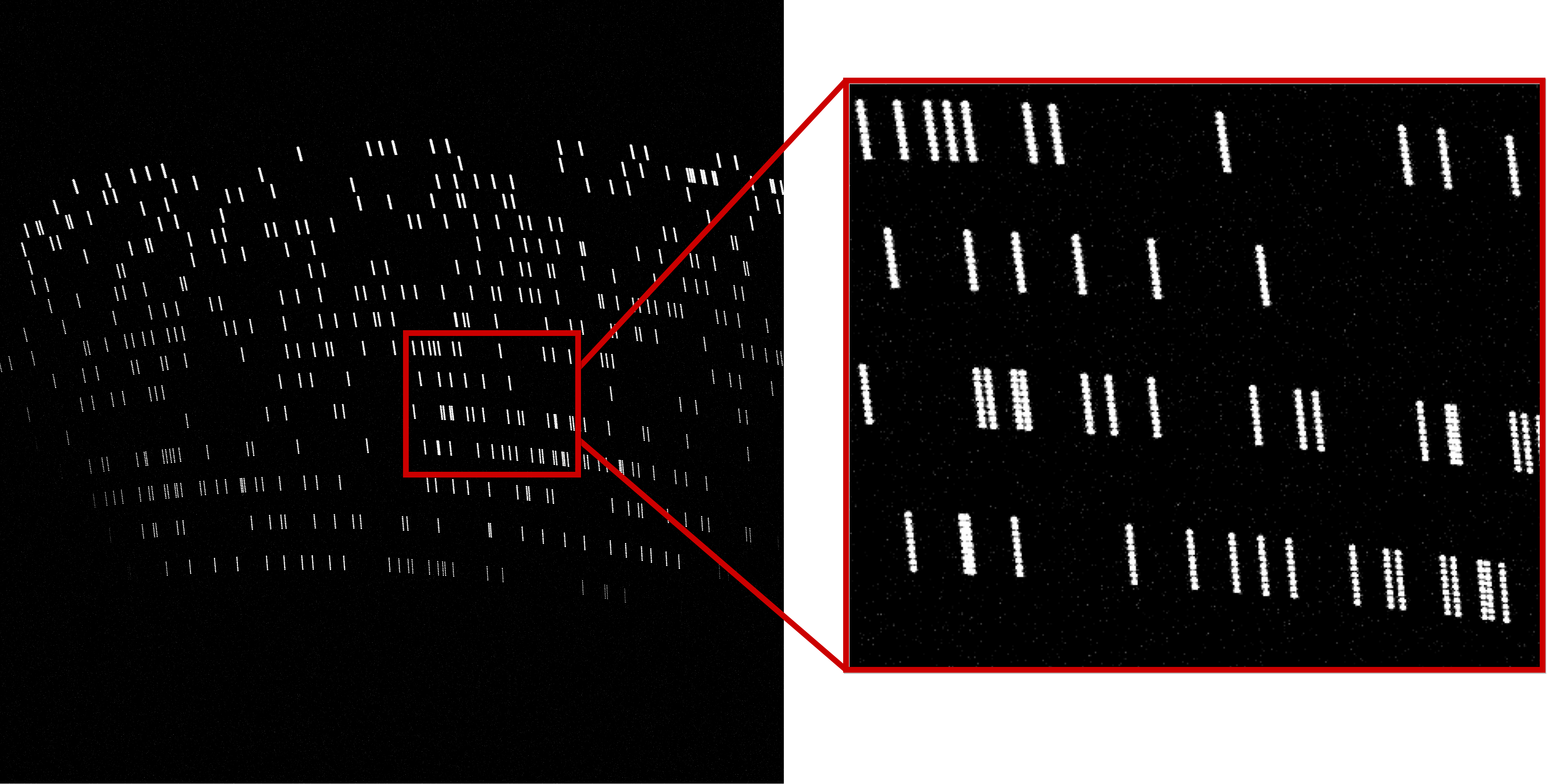}
\end{tabular}
\end{center}
\caption[example] 
%>>>> use \label inside caption to get Fig. number with \ref{}
{ \label{fig:e2e_calib_mph} 
NIR wavelength calibration frames with PenRays (Ar-Hg-Ne-Xe) emission lines and multi-pinhole mask.}
\end{figure}

%\newpage
An example of how a UV-VIS science frame would looks like can be seen in Fig. \ref{fig:uvvis_frame_1}. An observation of a G0V star, with V-mag = 18 (Vega System). The observation has been simulated again in new moon, at airmass=1.2 with sky PWV = 10 mm. The seeing assumed is 0.87 arcsec, at 550  nanometer , and the selected slit width is 1 arcsec. A single exposure of 1200 sec is emulated. The different quasi-orders traces with the object centered in the slit and sky emission features are clearly visible in the top and bottom quasi-orders, which are $r$-band and $i$-band respectively. While in the $u$-band (the second quasi-order from the top), the combined effect of star SED, instrument resolution, seeing and atmospheric transmission result in a trace that has 80\% lower photon flux with respect to the $i$-band for example. The slit tilt of the different spectral resolution elements images can be visually noticed mostly in the bottom quasi-order ($i$-band) thanks to greater contrast and presence of sky emission lines. The presence of hot/bad pixels and cosmic rays, for this frame, has been uniformly simulated across detector area.

\begin{figure} [h!]
\begin{center}
\begin{tabular}{c} %% tabular useful for creating an array of images 
\includegraphics[width=0.8\linewidth]{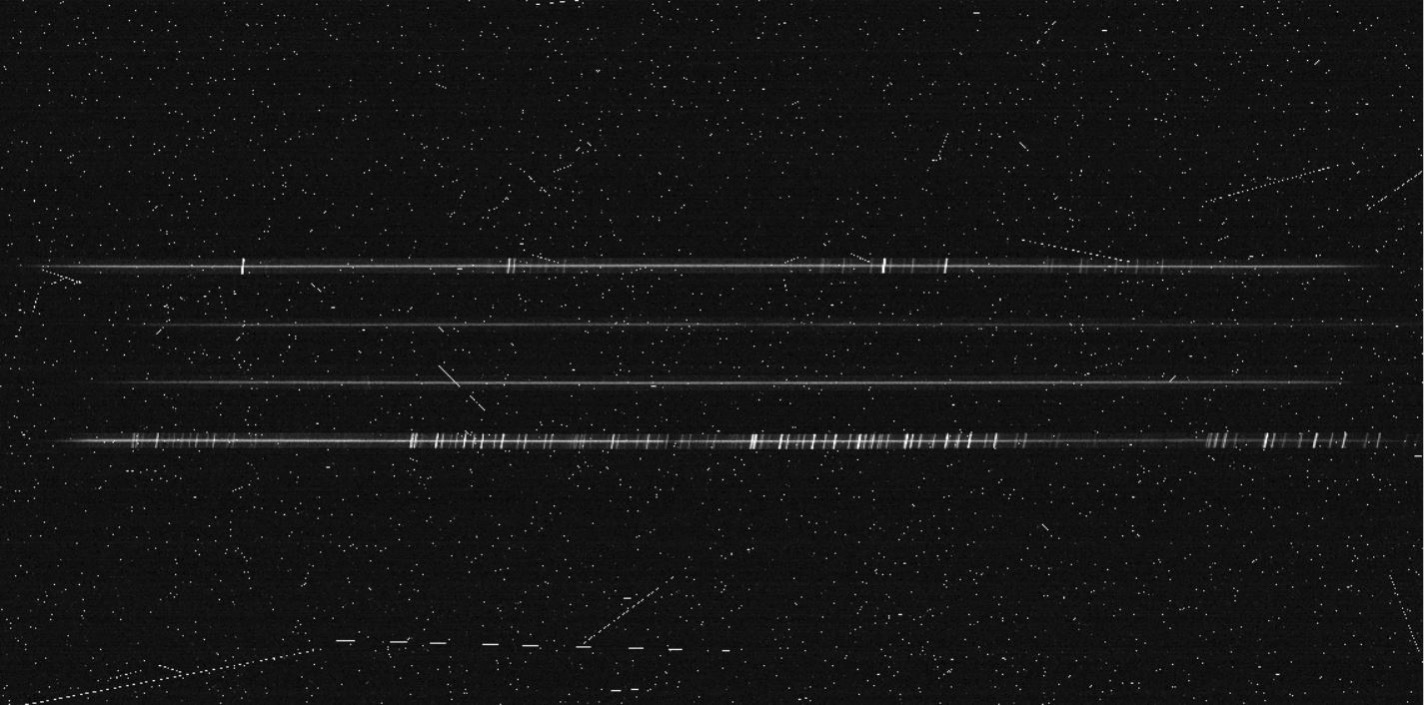}
\end{tabular}
\end{center}
\caption[example] 
%>>>> use \label inside caption to get Fig. number with \ref{}
{ \label{fig:uvvis_frame_1} 
UV-VIS arm simulated synthetic frame of a science observation of a G0V star $V$-mag = 18 (Vega System). See text for other simulation details.}
\end{figure}

Fig. \ref{fig:uvvis_frame_2} shows a flat slit calibration frame in the UV-VIS, with QTH+D2 combined sources and 10 sec of exposure. The $r$-band and $g$-band (1st and 3rd from top respectively) have a sharp cut-off at the spectral range edge. The $u$-band exhibits a lower flux at the beginning of its wavelength range due to the CCD-QE (see Fig. \ref{fig:qe}, left panel), while the $i$-band is affected by the UV-VIS/NIR dichroic transition region above 800  nanometer  (see Fig. \ref{fig:soxs_cp_vis_unit}), as can be seen from the right side of its trace (the bottom one).

\begin{figure} [h!]
\begin{center}
\begin{tabular}{c} %% tabular useful for creating an array of images 
\includegraphics[width=0.8\linewidth]{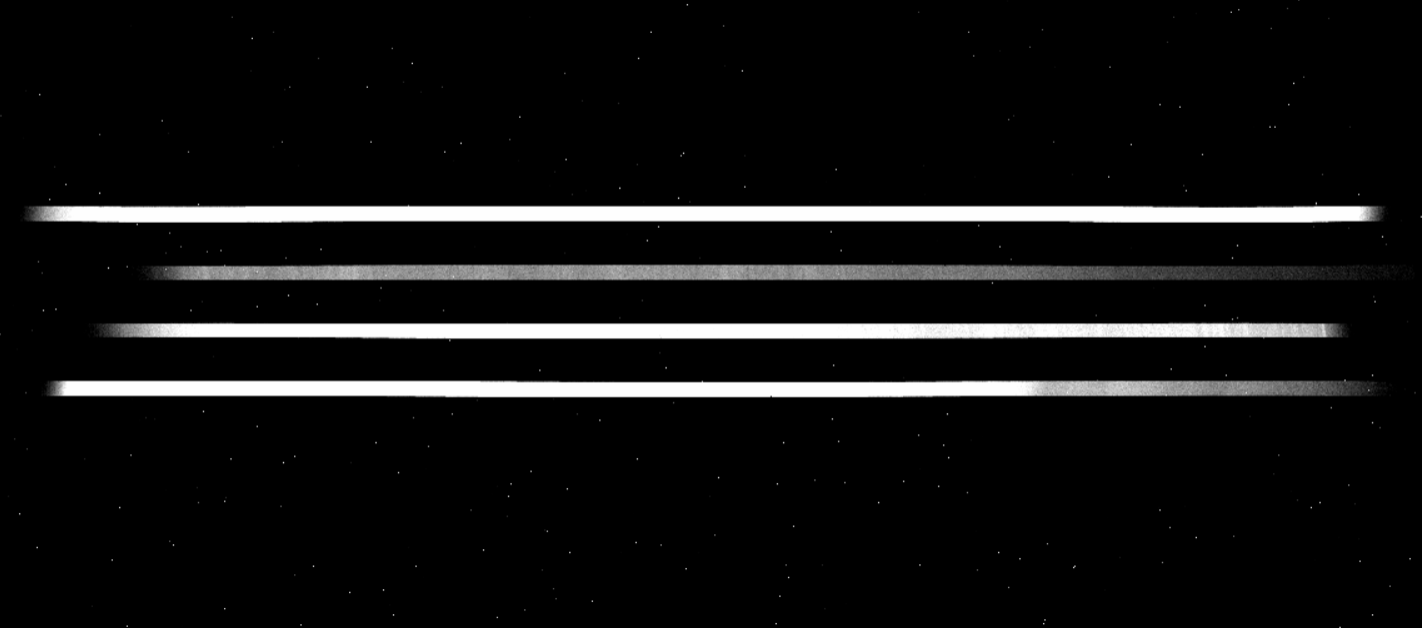}
\end{tabular}
\end{center}
\caption[example] 
%>>>> use \label inside caption to get Fig. number with \ref{}
{ \label{fig:uvvis_frame_2} 
UV-VIS calibration frame. Flat slit with QTH+D2 lamps and 10 sec of exposure.}
\end{figure}
\newpage
\subsection{Data Reduction Software Iteration}
\label{sec:e2e_drs}
It is a significant boon for the pipeline development team, and the SOXS Consortium as a whole, having simulated frames in the expected SOXS format before integration and commissioning of the instrument. 
Much of the pipeline development effort and debugging has already been performed ahead of time before the pipeline has been introduced to real SOXS data.

The SOXS data reduction pipeline (\texttt{soxspipe}, \cite{soxs_drs022}) development team have exploited the availability of simulated NIR SOXS data frames to test and hone many of the routines used across the reduction cascade. 

Fig. \ref{fig:soxspipe-one} show the results of a first-guess dispersion solution obtained by \texttt{soxspipe} using a simulated arc-lamp frame obscured by a single-pinhole mask. 
The arc lines detected in the frame (top image panel) are used to fit a global dispersion solution (second image panel). The residuals of the fit compared to the observed arc-line positions are presented in a x-y dispersion (residuals mean 0.13 pixels, standard deviation 0.07 pixels) and histogram plots in the bottom image panel.
Fig. \ref{fig:soxspipe-two} reveals how \texttt{soxspipe} then uses this first-guess dispersion solution, alongside a simulated flat-lamp frame, again obscured by the single-pinhole mask, to determine the traces of the centre of each of the spectral orders. 
Locations along the order centre traces (top image panel) are used to fit a global solution to the traces (second image panel). The residuals of the fit (residuals mean 0.04 pixels, standard deviation 0.03 pixels) and a plot of the trace FWHM (along spatial direction, in units of pixels) dependence on wavelength can be found in the bottom panels.
Fig. \ref{fig:soxspipe-three} shows the use of the centre trace solutions and a set of simulated NIR slit lamp-flat frames to a) create a master-flat frame and b) detect and fit the edges of each order on the frame. 
Measured pixel locations of the order edges (top image panel) are used to fit a global solution to the edges (second image panel). The residuals of the fit compared to the measured edge pixel locations can be found in the bottom panels (residuals mean 0.12 pixels, standard deviation 0.09 pixels).
Finally, Fig. \ref{fig:soxspipe-four} shows the use of the first-guess dispersion solution, the order location solutions and a simulated arc-frame, obscured this time by a multi-pinhole mask, to determine the final full-dispersion and spatial solution required to reduce object frames later in the reduction cascade.
The arc lines detected in the frame (top image panel) are used to fit a global dispersion solution (second image panel). The residuals of the fit compared to the observed arc-line positions can be found in the bottom panels as x-y dispersion (residuals mean 0.15 pixels, standard deviation 0.09 pixels) and histogram plots.

Alongside simulated SOXS data, \texttt{soxspipe} has the ability to reduce X-shooter data. The low residuals achieved while finding global best-fit polynomials for the order-centre traces, order-edges and dispersion solutions within simulated SOXS data, closely resemble those found reducing X-shooter data with both \texttt{soxspipe} and the original X-shooter data-reduction pipeline \cite{2011AN....332..227G}. 

\begin{figure} [h!]
\begin{center}
\begin{tabular}{c} %% tabular useful for creating an array of images 
\includegraphics[width=0.9\linewidth]{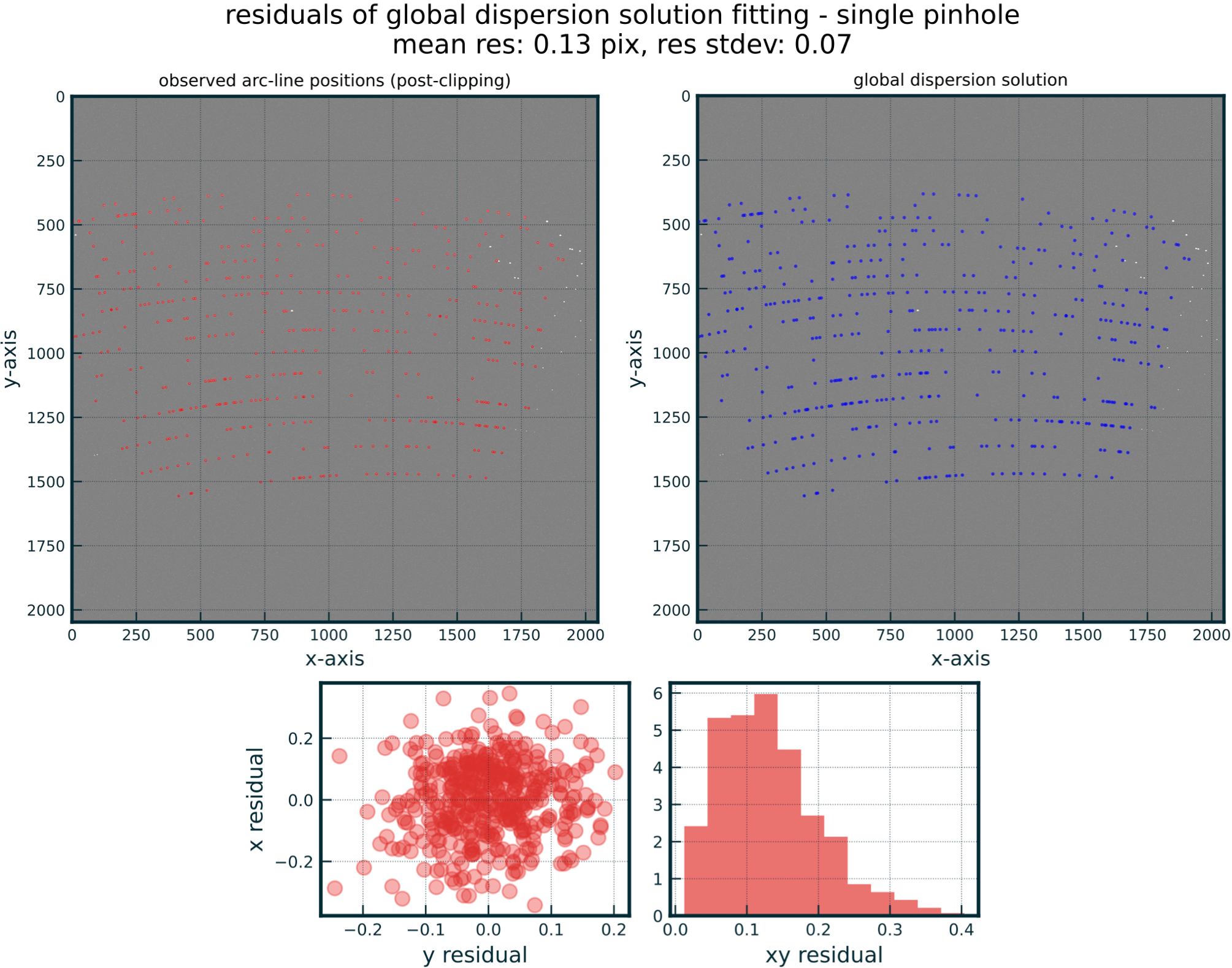}
\end{tabular}
\end{center}
\caption[example] 
{\label{fig:soxspipe-one} 
The resulting first-guess dispersion solution and residuals as fitted by the SOXS data-reduction pipeline using a simulated arc-lamp frame obscured by a single-pinhole mask.} 
%The arc lines detected in the frame (top image panel) are used to fit a global dispersion solution (second image panel). The residuals of the fit compared to the observed arc-line positions and a plot of the trace FWHM (in pixels) dependence on wavelength can be found in the bottom panels.}
\end{figure}

\begin{figure} [h!]
\begin{center}
\begin{tabular}{c} %% tabular useful for creating an array of images 
\includegraphics[width=0.9\linewidth]{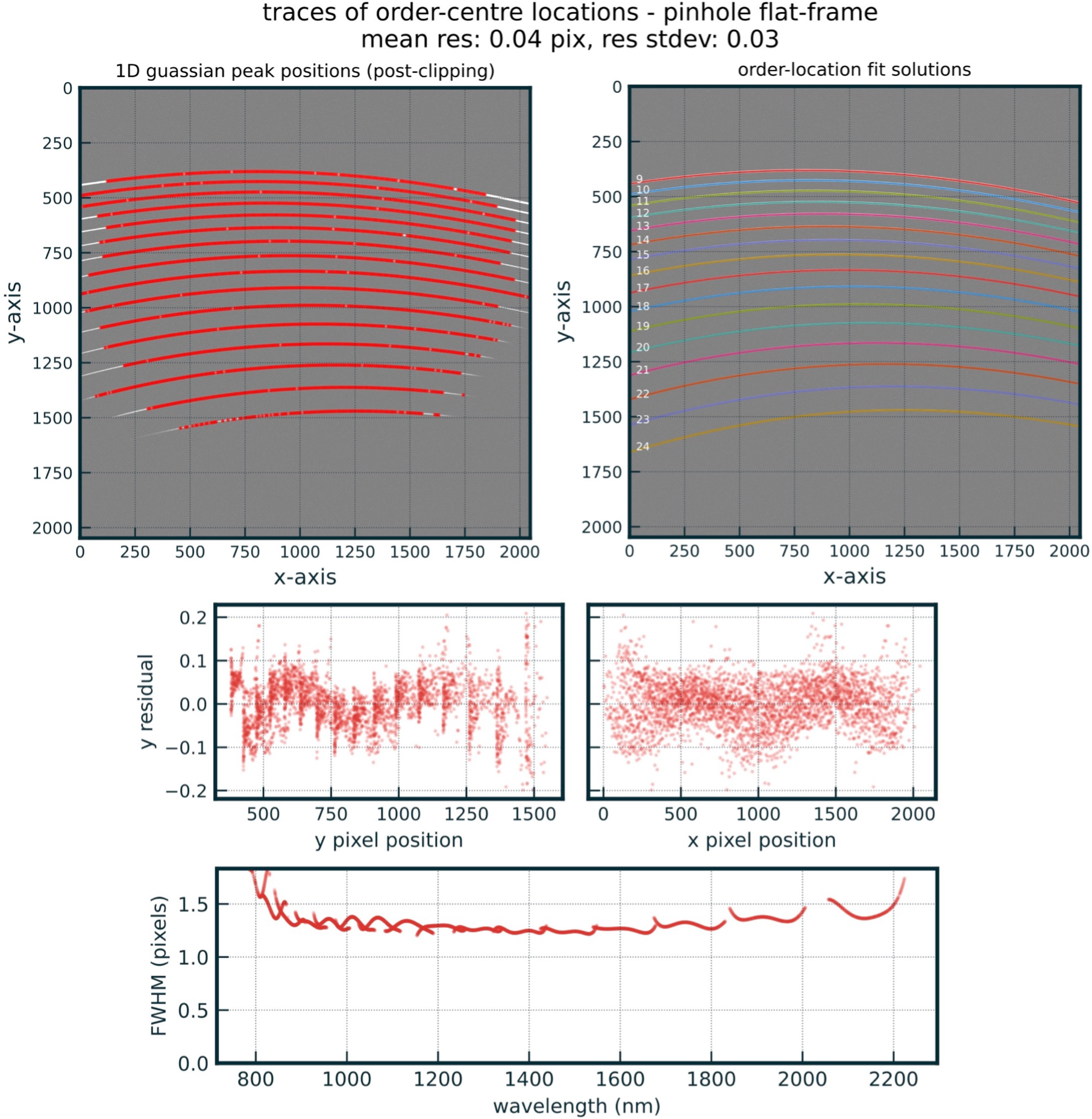}
\end{tabular}
\end{center}
\caption[example] 
{\label{fig:soxspipe-two} 
The order-centre traces fitted by the SOXS data-reduction pipeline using a simulated flat-lamp frame obscured by a single-pinhole mask.} %Locations along the order centre traces (top image panel) are used to fit a global solution to the traces (second image panel). The residuals of the fit compared to the observed trace pixel locations and a plot of the trace FWHM (in pixels) dependence on wavelength can be found in the bottom panels.}
\end{figure}

%\clearpage
\begin{figure} [h!]
\begin{center}
\begin{tabular}{c} %% tabular useful for creating an array of images
\includegraphics[width=0.9\linewidth]{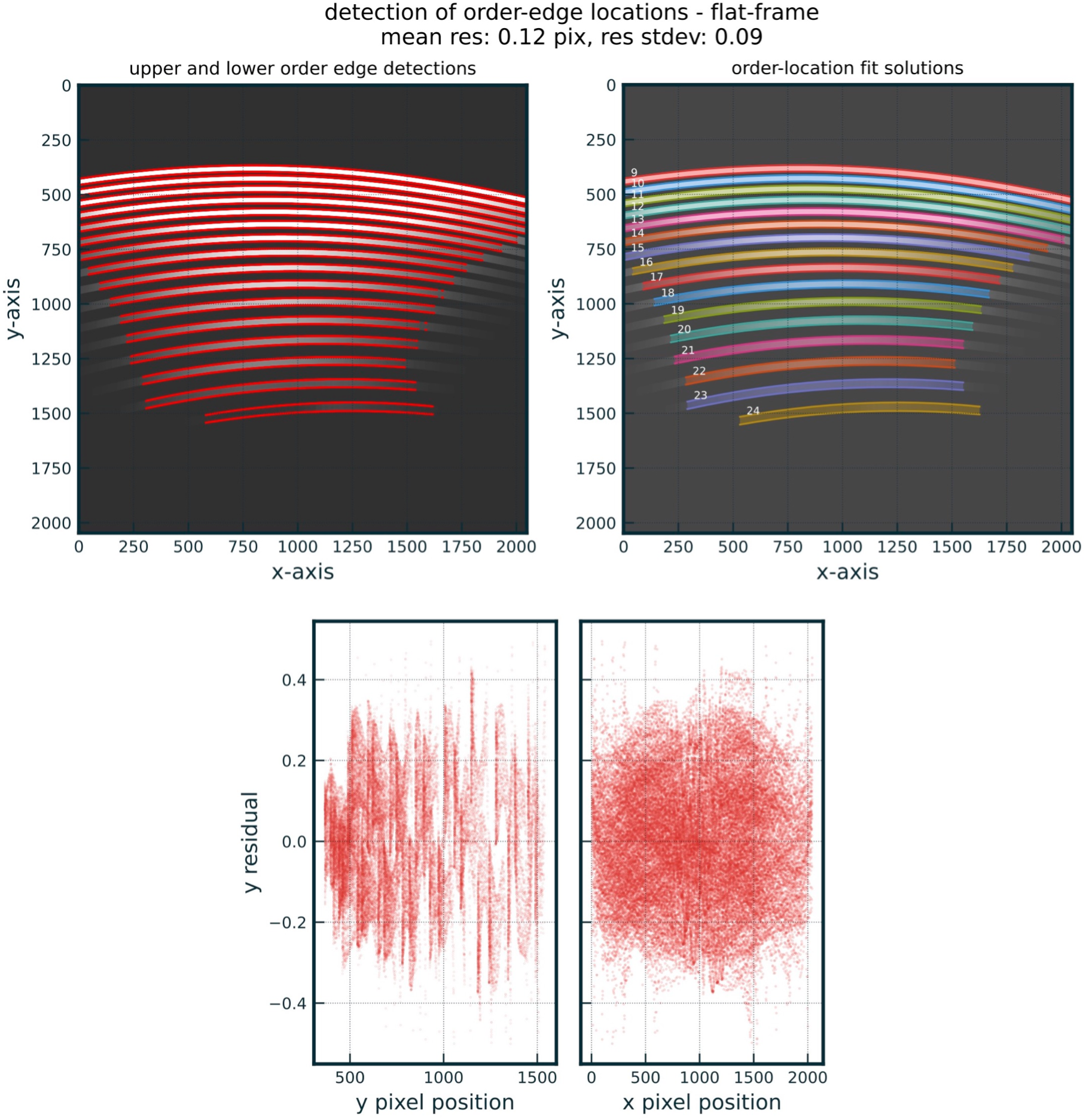}
\end{tabular}
\end{center}
\caption[example] 
{\label{fig:soxspipe-three} 
The order-edges fitted by the SOXS data-reduction pipeline using a master-flat frame created from a set of simulated full slit flat-lamp frames.}
%Measured pixel locations of the order edges (top image panel) are used to fit a global solution to the edges (second image panel). The residuals of the fit compared to the measured edge pixel locations can be found in the bottom panels.}
\end{figure}

%\clearpage
\begin{figure} [h!]
\begin{center}
\begin{tabular}{c} %% tabular useful for creating an array of images 
\includegraphics[width=0.9\linewidth]{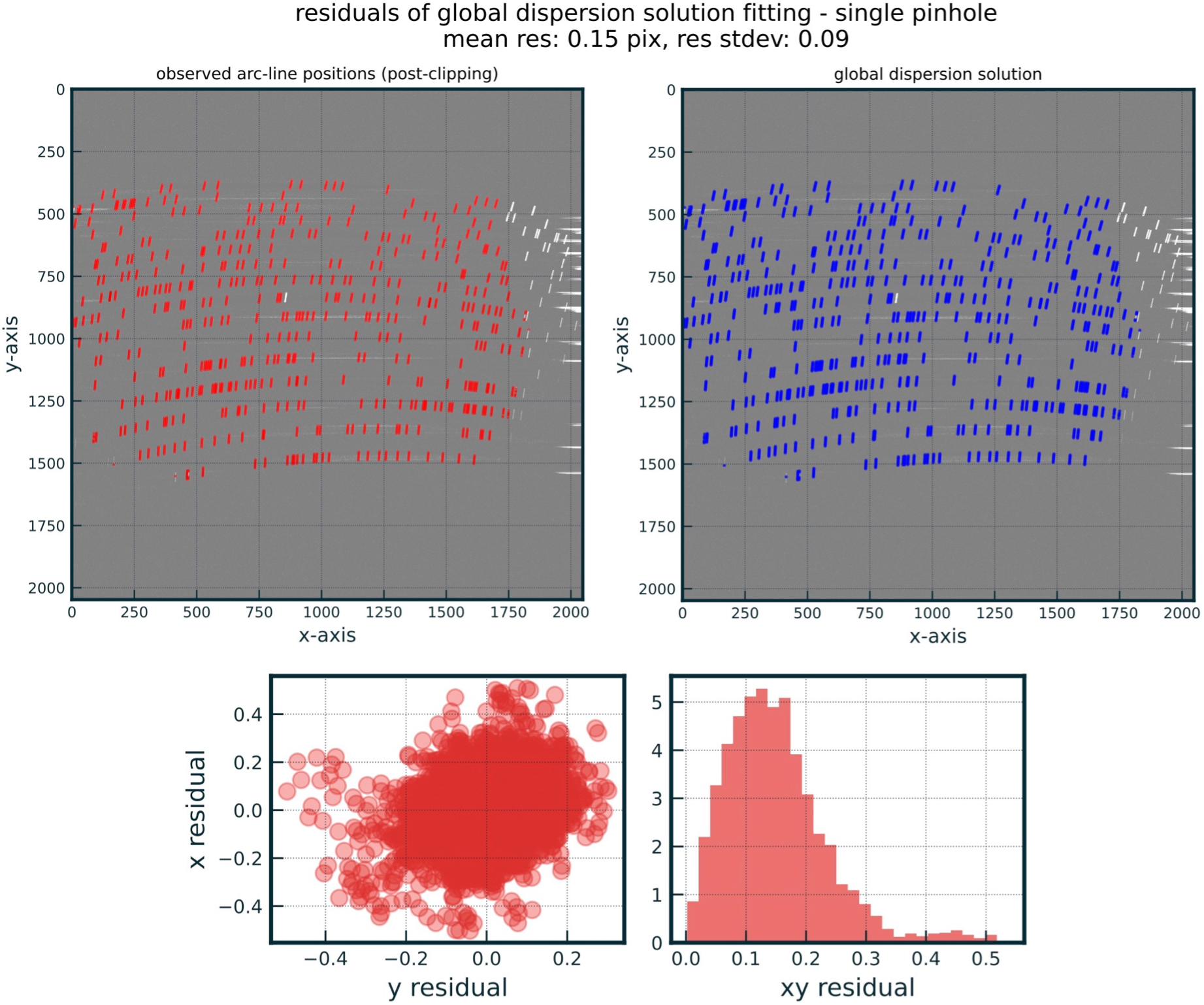}
\end{tabular}
\end{center}
\caption[example] 
{\label{fig:soxspipe-four} 
The resulting final dispersion solution and residuals as fitted by the SOXS data-reduction pipeline using a simulated arc-lamp frame obscured by a multi-pinhole mask.}
%The arc lines detected in the frame (top image panel) are used to fit a global dispersion solution (second image panel). The residuals of the fit compared to the observed arc-line positions can be found in the bottom panels.}
\end{figure}

%\newpage
\section{Exposure Time Calculator}
\label{sec:etc}
The Exposure Time Calculator of SOXS is a web-based tool that reflects the two different modes of observation, spectroscopy and imaging. 
The architecture for both modes was developed based on the structure of the previously described end-to-end simulator, sharing at a high level a similar workflow of modules and units and at lower level the required mathematical recipes. 
Each unit was tested individually and compared with several ETCs from ESO. 
The ETC was developed via Python 3.9 using the web application framework Flask (2.0). 
This tool has already been useful at different stages of development to evaluate the performance of the instrument (see comparison with VLT-X-shooter for Spectroscopy and with NTT-EFOSC2 for imaging in Section \ref{sec:etc_output}) and it will be widely used by scientists when SOXS will be in operations. 

\subsection{ETC Architecture}
\label{sec:etc_architecture}
The workflow of the ETC starts from generating a synthetic 1D spectrum in units of  photon s cm$^2$ angstrom and computing the Sky Radiance and the Atmospheric Transmission profiles (from ESO's SkyCalc tool) as already described in Sections \ref{sec:e2e_architecture_sci_module} and \ref{sec:e2e_architecture_sky_module}). 
%The sky radiance spectrum loaded is in units of \unit{\photon/\s/\meter^2/\micro\meter/\arcsec^2}, thus it is first calculated for the on-sky area related to the slit length and selected slit width according to the simulation (both in arcsec) and then the units are converted in \unit{\photon/\s/\cm^2/\angstrom}. 
The number of object and sky counts in terms of photo-electrons per spectral bin are then be calculated by integrating over a delta lambda (wavelength bin) per SRE or per Pixel based on user selection, multiplying by the telescope effective area and the total exposure time (according to the single exposure integration time and number of exposures, see Fig. \ref{fig:etc_ins}).  
The final results, provided for both UV-VIS and NIR arms, take into account the different optical and slit efficiencies and the detector noises.

\subsection{Layout}
\label{sec:etc_layout}
The design of the web application was developed to be intuitive and easy to use; once the spectroscopy or imaging mode is initially selected via a toggle, the user has to fill in the several parameters, relevant for the observations, through a public HTML interface via the SOXS website\footnote{\url{http://www.brera.inaf.it/~campana/SOXS/Son\_of\_X-Shooter.html}} as described in the following. 
The 1D synthetic spectrum can be obtained from four different selection's of SEDs:
\begin{itemize}
  \item Black Body;
  \item Power-Law;
  \item User-defined Spectrum (Possibility of selecting a pre-loaded spectrum);
  \item Single Emission Line.
 \end{itemize}
The user must also select the magnitude of the object with its reference band and system, the redshift and if the spatial distribution is point-like or from an extended source, as shown in Fig. \ref{fig:etc_sed}. 
The different parameters box are in light grey or dark grey according to the type of selected SED. For example, if power law is selected, only the power-law index and magnitude parameters can be set, while the others cannot be modified.
Subsequently, the call to ESO's SkyCalc tool will be made by setting the following inputs (see box in in Fig. \ref{fig:etc_sky}):
 \begin{itemize}
  \item Days from new moon;
  \item Airmass;
  \item Precipitable Water Vapor (PWV);
  \item Seeing (reference value at zenith and 500  nanometer ).
 \end{itemize}
%\newpage

\begin{figure} [ht]
\begin{center}
\begin{tabular}{c} %% tabular useful for creating an array of images™
\includegraphics[width=0.872\linewidth]{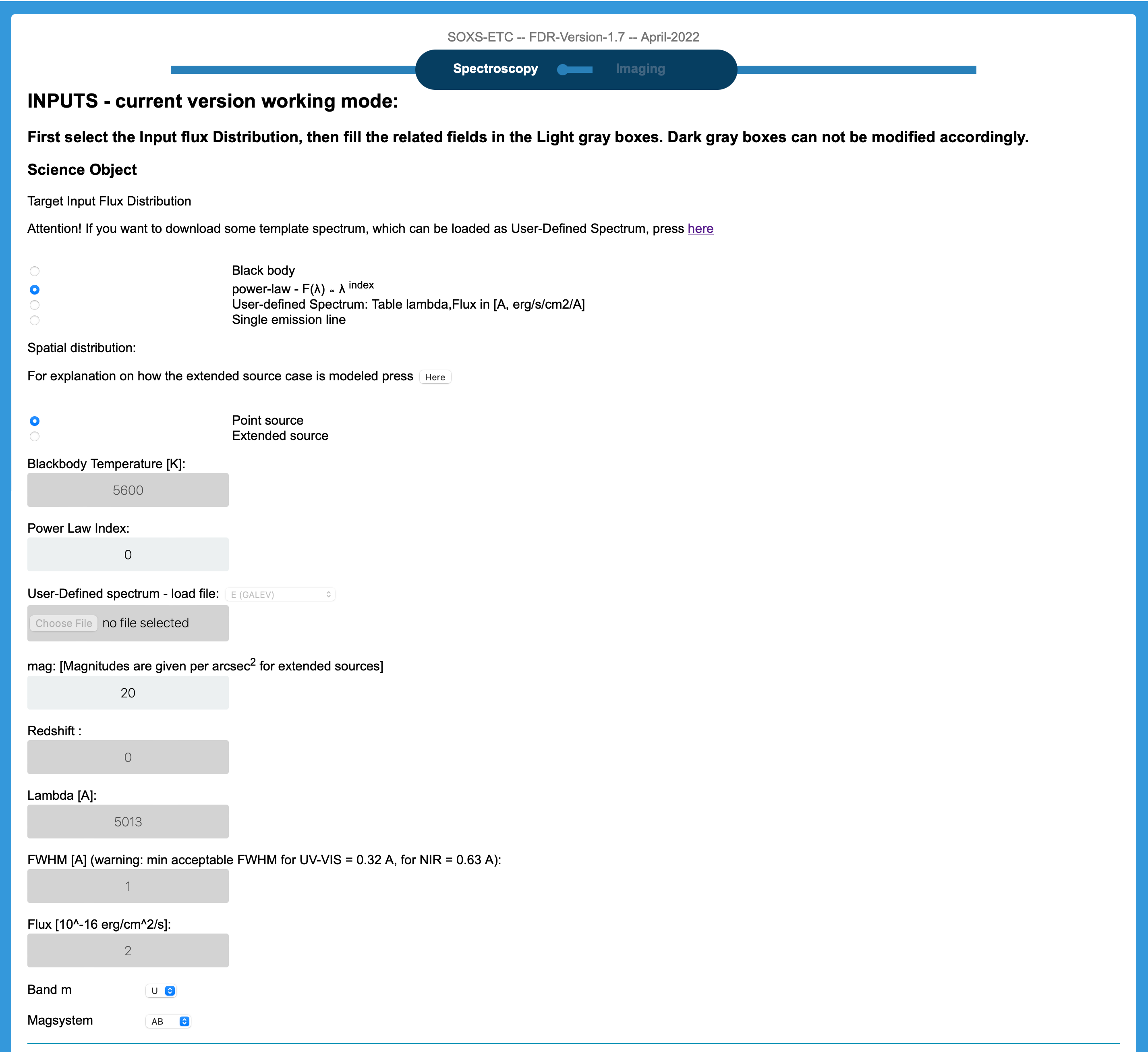}
\end{tabular}
\end{center}
\caption[example] 
%>>>> use \label inside caption to get Fig. number with \ref{}
{ \label{fig:etc_sed} 
Graphical layout of the ETC simulator of SOXS. Science-Object input flux distribution.
}
\end{figure}
\begin{figure} [ht]
\begin{center}
\begin{tabular}{c} %% tabular useful for creating an array of images™
\includegraphics[width=0.872\linewidth]{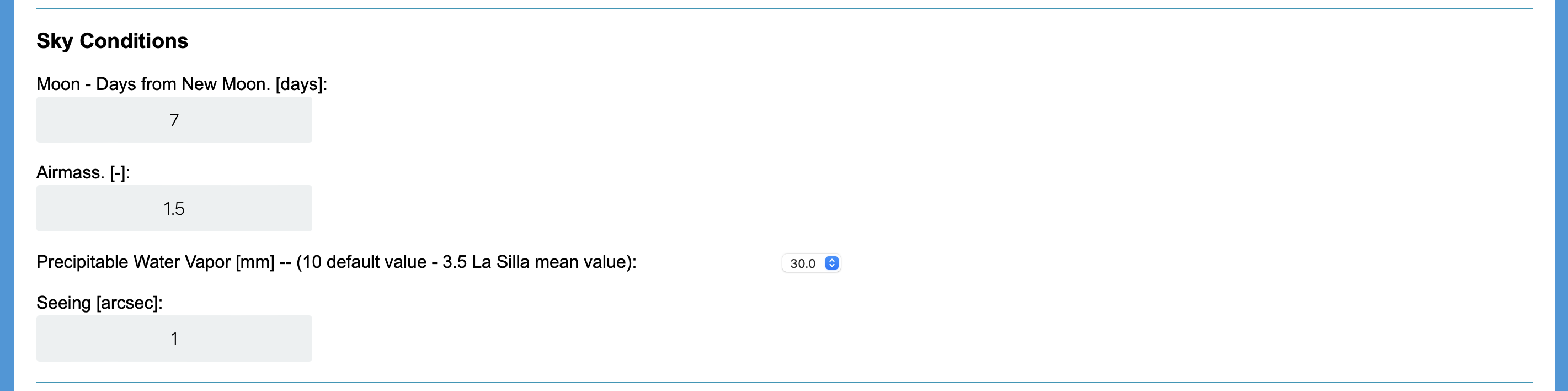}
\end{tabular}
\end{center}
\caption[example] 
%>>>> use \label inside caption to get Fig. number with \ref{}
{ \label{fig:etc_sky} 
Graphical layout of the ETC simulator of SOXS. Sky and Atmospheric condition.
}
\end{figure}
\newpage

Finally, in spectroscopy mode (Fig. \ref{fig:etc_ins}), the instrument set-up consists of the parameters of exposure time and slit size, that can be selected independently between the two arms. In addition, different number of Spectral Resolution Element per order and binning modes (only for the UV-VIS arm detector) can be selected.

While for imaging mode, a filter (among the following: {\it u, g, r, i, z, y} (LSST) and $V$ Johnson bands) and the total exposure time (exposure time for single exposure and number of exposures) parameters can be selected.

\begin{figure} [ht]
\begin{center}
\begin{tabular}{c} %% tabular useful for creating an array of images™
\includegraphics[width=0.872\linewidth]{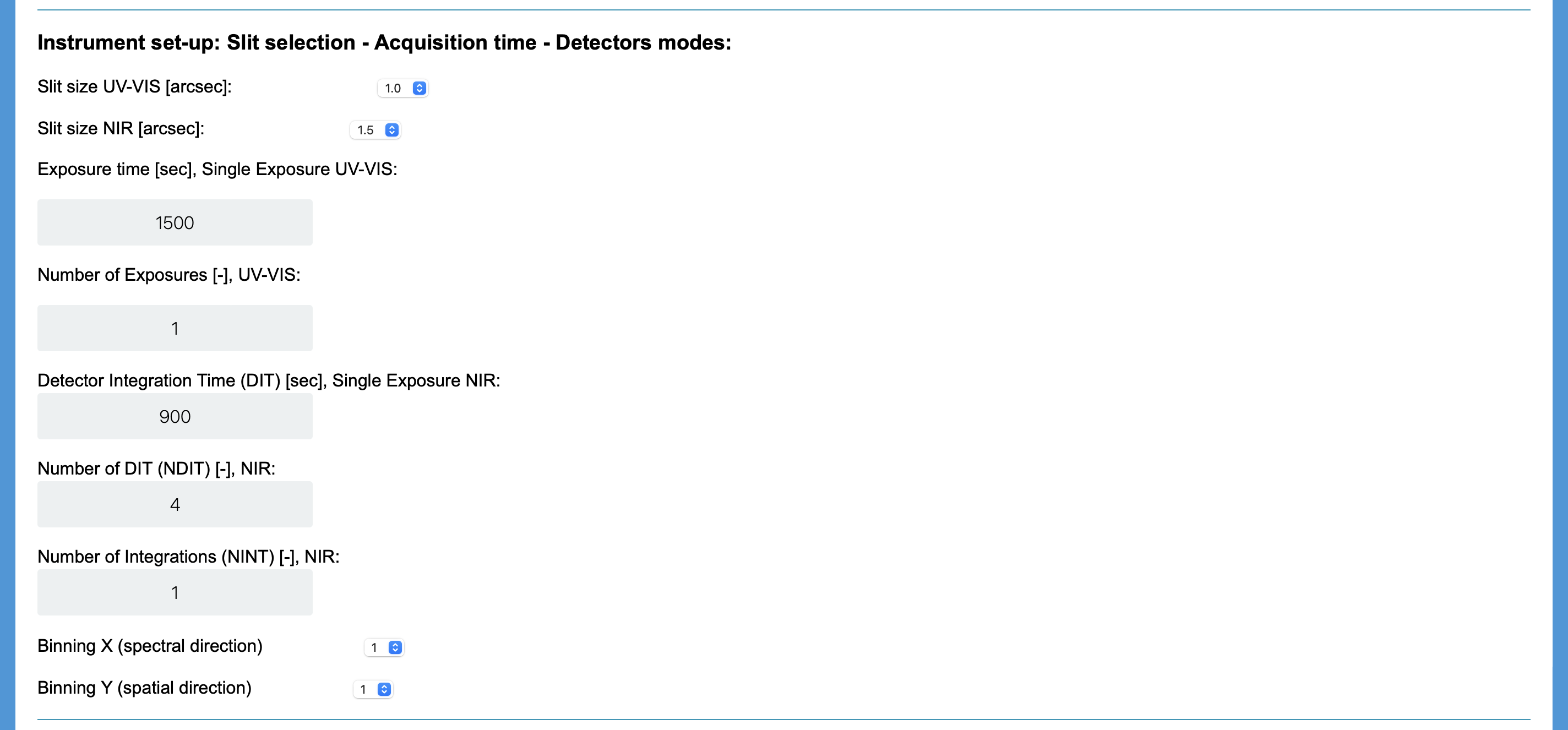}
\end{tabular}
\end{center}
\caption[example] 
%>>>> use \label inside caption to get Fig. number with \ref{}
{ \label{fig:etc_ins} 
Graphical layout of the ETC simulator of SOXS. Instrument set-up input parameters.}
\end{figure}

\newpage
\subsection{Output Results}
\label{sec:etc_output}
The ETC provides different outputs as listed below and in the case of spectroscopy, the same kind of information will be provided for both the UV-VIS and NIR arms.
\begin{itemize}
  \item SNR (Spectroscopy \& Imaging);
  \item Total efficiency (Spectroscopy \& Imaging) and Filter efficiency (Imaging);
  \item Source Spectrum (Spectroscopy \& Imaging);
  \item Sky radiance Spectrum \& Atmospheric Transmission (Spectroscopy \& Imaging);
  \item Source \& Sky Photoelectrons (Spectroscopy \& Imaging);
  \item Noises (Spectroscopy \& Imaging);
  \item Maximum Intensity (Spectroscopy).
 \end{itemize}

In the following example, a comparison of the output SNR between SOXS and X-shooter is presented, using both ETCs (see Fig. \ref{fig:soxs_uvvis_nir_plot} and Fig. \ref{fig:xs_uv_vis_nir_plot}).
This is done simulating a 5600 K Black Body with $R$-mag = 21.5 (AB System) observed with 4 exposures for 900 seconds and the following sky-condition: Moon phase = 0 (i.e. new moon), air-mass = 1.2, precipitable water vapor (PWV) = 30  mm  and seeing = $1"$ (at 5000  Angstrom ).
In the X-shooter-VIS band SOXS has a higher SNR because in this range, although the NTT telescope effective area is more than four times smaller than the VLT one, its peak instrumental efficiency is about 30\% higher than X-shooter and its dispersion is such that the wavelength coverage per pixel is approximately four times larger.
In the X-shooter-UVB band, SOXS has a similar instrumental efficiency but dispersion in this region is not balancing the larger effective area of the VLT.
In the NIR arm, being SOXS extremely similar to X-shooter, the better performance of the latter is due to the advantages of a larger effective telescope area and telescope throughput.

\begin{figure} [ht]
\begin{center}
\begin{tabular}{c} %% tabular useful for creating an array of images™
\includegraphics[height=5cm]{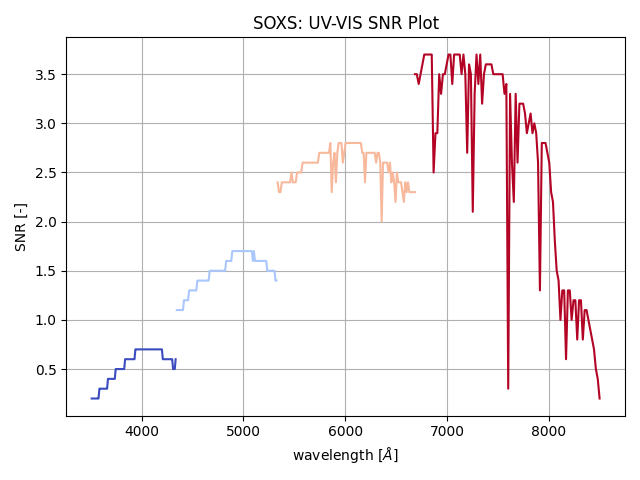}
\includegraphics[height=5cm]{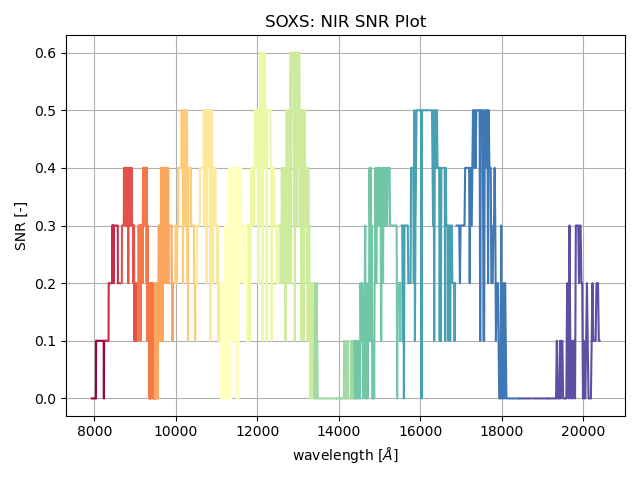}
\end{tabular}
\end{center}
\caption[example] 
%>>>> use \label inside caption to get Fig. number with \ref{}
{ \label{fig:soxs_uvvis_nir_plot} 
Black Body: 5600 K with magnitude of 21.5 (AB) $R$-Band observed with 4 exposures for 900 seconds.
}
\end{figure}

\begin{figure}[htp]
\begin{center}
\subfloat[UVB]{\label{fig:a}\includegraphics[width=0.33\linewidth]{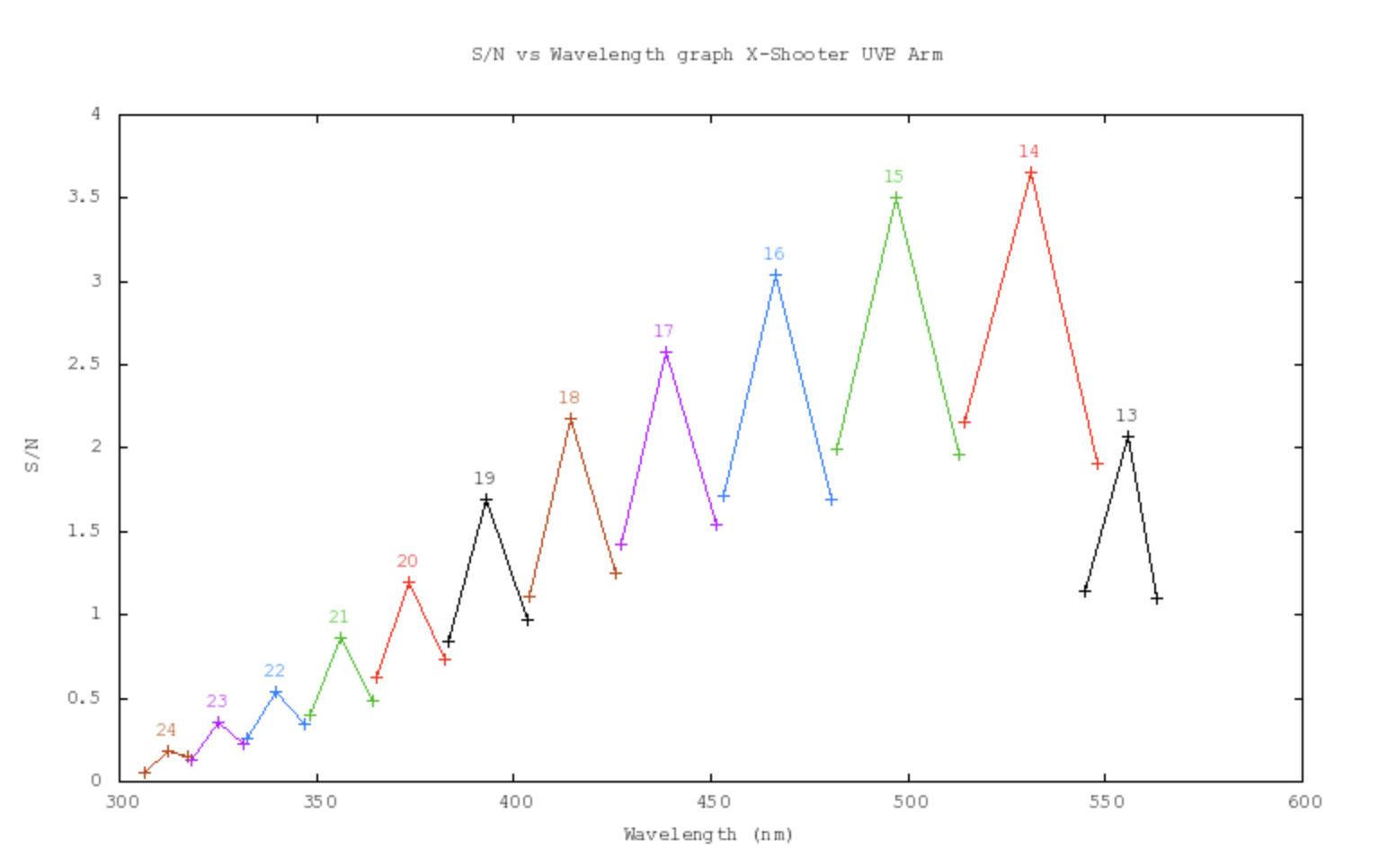}}
\subfloat[VIS]{\label{fig:b}\includegraphics[width=0.33\linewidth]{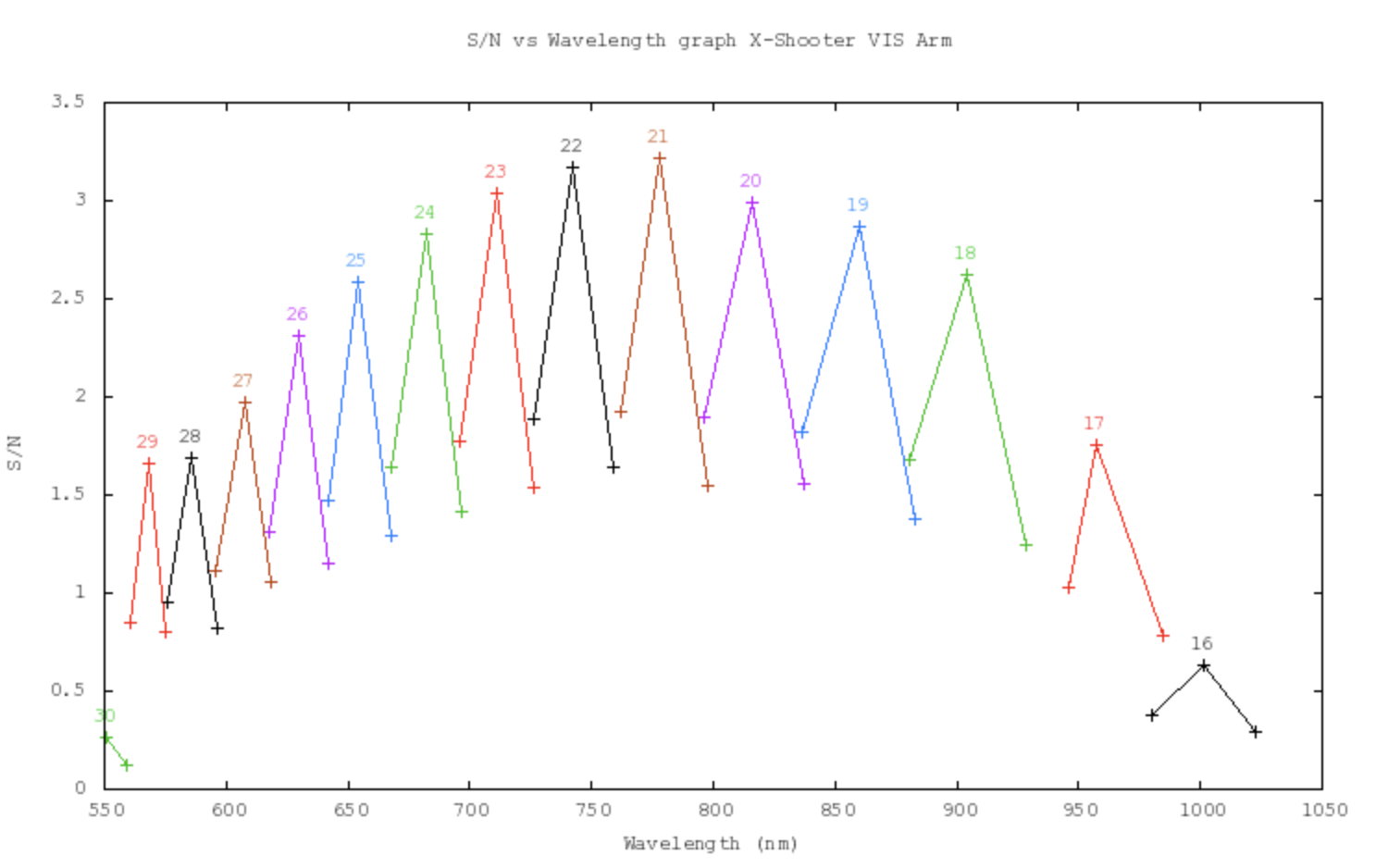}}
\subfloat[NIR]{\label{fig:c}\includegraphics[width=0.33\linewidth]{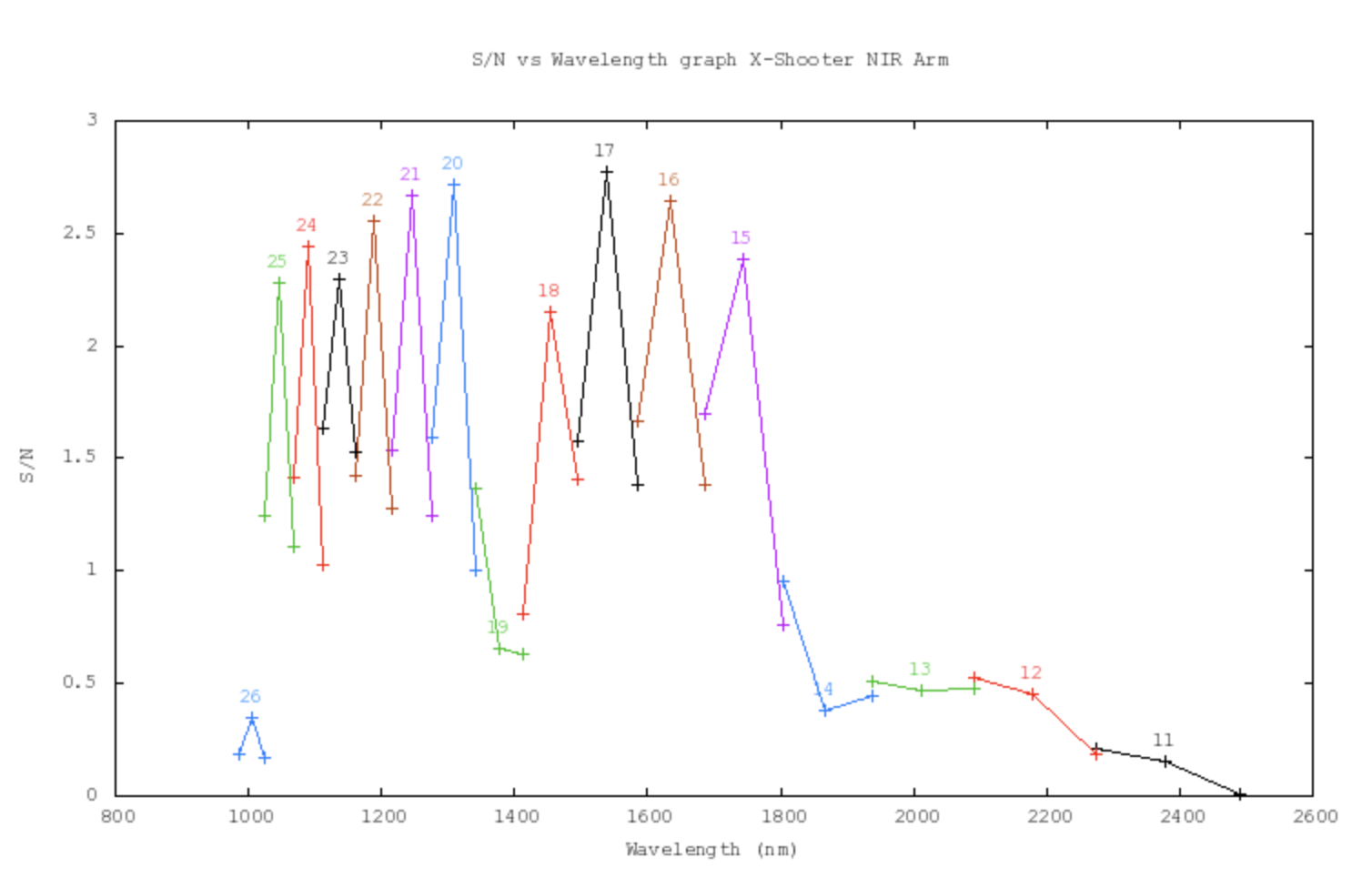}}
\caption[example] 
%>>>> use \label inside caption to get Fig. number with \ref{}
{ \label{fig:xs_uv_vis_nir_plot} 
Black Body: 5600 K with magnitude of 21.5 (AB) $R$-Band observed with 4 exposures of 900 seconds.
}
\end{center}
\end{figure}
%\newpage

Regarding the imaging output, the following results were compared with the EFOSC2 instrument and can be found in Table \ref{tab:acq_soxs_efosc2}; showing that by simulating a Black Body we obtain for the blue-red filters ({\it V, g, r}) a slightly worse outcome, while for the reddest filters ({\it i, z}) the performance is better.
%BB=5600K, mag R AB=24.5
%Single exposure 1500s, 0d Moon, 1.2 airmass, 1' seeing
\begin{table}[h!]
\caption{Performances of the ACQ Camera of SOXS comparable to EFOSC2 simulating a} 
\label{tab:acq_soxs_efosc2}
\begin{center}       
\begin{tabular}{|l|l|l|}
\hline
\rule[-1ex]{0pt}{3.5ex}  SNR & SOXS & EFOSC2  \\
\hline
\rule[-1ex]{0pt}{3.5ex}  $V$ & 7.7 & 10   \\
\hline
\rule[-1ex]{0pt}{3.5ex}  $g$ & 9.4 & 9.8 \\
\hline
\rule[-1ex]{0pt}{3.5ex}  $r$ & 10.18 & 10  \\
\hline
\rule[-1ex]{0pt}{3.5ex}  $i$ & 7.3 & 6.1  \\
\hline
\rule[-1ex]{0pt}{3.5ex}  $z$ & 4.3 & 3.1 \\
\hline 
\end{tabular}
\end{center}
\end{table}
\newpage
\section{Conclusion}
\label{sec:conclusion}
In this contribution we first described the architecture and development of the NTT-SOXS E2E instrument simulator, showing both the different modules tasks and functionalities and how the Image simulation Module produce a synthetic image for both the instrument arms. Examples of simulations regarding different frames and the results obtained from iterating with the DRS for the NIR arm have been presented. This cross-check has also been fundamental to debug the simulator and to validate a full end-to-end simulation. A specific set of calibration will also be generated and used to aid the UV-VIS arm. Finally, we described the Exposure Time Calculator along with its architecture, layout and multiple outputs, for both spectroscopy and imaging observation modes.
\newpage
% References
\bibliography{report} % bibliography data in report.bib

\begin{thebibliography}{10}

\bibitem{MOONS}
{Li Causi}, G. et~al., ``{Virtual MOONS: a focal plane simulator for the MOONS
  thousand-fiber NIR spectrograph},'' {\em Proc. of SPIE Vol.} {\bf 9147},
  914764 (Aug. 2014).

\bibitem{Radial_vel_error_budget}
Bechter, A.~J., Bechter, E.~B., Jr., J. R.~C., King, D., and Crass, J., ``{A
  radial velocity error budget for single-mode Doppler spectrographs},'' {\em
  Proc. of SPIE} {\bf 10702},  2075 -- 2099 (2018).

\bibitem{iLocater_e2e_drs}
Bechter, E.~B., Bechter, A.~J., Crepp, J.~R., Crass, J., and King, D.,
  ``Instrument simulator and data reduction pipeline for the {iLocater}
  spectrograph,'' {\em Publ. Astron. Soc. Pac.}~{\bf 131},  996 (jan 2019).

\bibitem{websim}
Puech, M., Yang, Y., Jégouzo, I., et~al., ``{WEBSIM-COMPASS: a new generation
  scientific instrument simulator for the E-ELT},'' {\em Proc. of SPIE} {\bf
  9908},  2919 -- 2933 (2016).

\bibitem{soxs_update}
Schipani, P., Campana, S., Claudi, R., Aliverti, M., et~al., ``Development
  status of the {SOXS} spectrograph for the {ESO}-{NTT} telescope,'' in [{\em
  Ground-based and Airborne Instrumentation for Astronomy
  {VIII}}{\nolinebreak\hspace{0.1em}]},  Evans, C.~J., Bryant, J.~J., and
  Motohara, K., eds., {SPIE} (dec 2020).

\bibitem{soxs_gen_2022}
Schipani, P. et~al., ``Progress on the {SOXS} transients chaser for the
  {ESO}-{NTT},'' {\em Proc. SPIE} {\bf 12184-24} (2022).

\bibitem{soxs_cp}
Claudi, R. et~al., ``{The common path of SOXS (Son of X-Shooter)},'' {\em Proc.
  SPIE} {\bf 10702},  1189 -- 1199 (2018).

\bibitem{acq_guid_soxs}
Brucalassi, A. et~al., ``{Final Design and development status of the
  Acquisition and Guiding System for {SOXS}},'' {\em Proc. SPIE} {\bf 114475V}
  (2020).

\bibitem{MITS}
Rubin, A. et~al., ``Mits: the multi-imaging transient spectrograph for soxs,''
  {\em Proc. SPIE} {\bf 10702} (2018).

\bibitem{soxs_vis}
Cosentino, R. et~al., ``The vis detector system of {SOXS},'' {\em Proc. SPIE}
  {\bf 10702} (2018).

\bibitem{soxs_optical_design}
S{\'{a}}nchez, R.~Z. et~al., ``Optical design of the {SOXS} spectrograph for
  {ESO} {NTT},'' {\em Proc. SPIE} {\bf 10702} (jul 2018).

\bibitem{soxs_nir_paper}
Vitali, F. et~al., ``{The NIR spectrograph for the new SOXS instrument at the
  NTT},'' {\em Proc. SPIE} {\bf 10702},  697 -- 709 (2018).

\bibitem{soxs_cal}
Kuncarayakti, H. et~al., ``Design and development of the {SOXS} calibration
  unit,'' {\em Proc. SPIE} {\bf 11447} (dec 2020).

\bibitem{soxs_cp_2022}
Radhakrishnan, K. et~al., ``From assembly to the complete integration and
  verification of the {SOXS} common path,'' {\em Proc. SPIE} {\bf 12184-305}
  (2022).

\bibitem{soxs_acq_2022}
Araiza-Dur{\'{a}}n, J.~A. et~al., ``The integration and alignment phase for the
  acquisition and guiding system of {SOXS},'' {\em Proc. SPIE} {\bf 12184-306}
  (2022).

\bibitem{soxs_nir_2022}
Vitali, F. et~al., ``Progress on the {SOXS} {NIR} spectrograph {AIT},'' {\em
  Proc. SPIE} {\bf 12184-302} (2022).

\bibitem{soxs_e2e_2020}
Genoni, M., Landoni, M., Causi, G.~L., et~al., ``{SOXS} end-to-end simulator:
  development and applications for pipeline design.,'' {\em Proc. SPIE} {\bf
  114501B} (2020).

\bibitem{skycalc}
ESO, ``Skycalc,'' (2013).
\newblock
  \url{http://www.eso.org/observing/etc/bin/gen/form?INS.MODE=swspectr+INS.NAME=SKYCALC}.

\bibitem{numba}
Lam, S.~K., Pitrou, A., and Seibert, S., ``Numba: A llvm-based python jit
  compiler,'' in [{\em Proceedings of the Second Workshop on the LLVM Compiler
  Infrastructure in HPC}{\nolinebreak\hspace{0.1em}]},  {\em LLVM '15},
  Association for Computing Machinery, New York, NY, USA (2015).

\bibitem{pyxel_2020}
Prod'homme, T., Lemmel, F., Arko, M., et~al., ``Pyxel: the collaborative
  detection simulation framework,'' {\em Proc. SPIE} {\bf 1145408} (2020).

\bibitem{2011AN....332..227G}
{Goldoni}, P., ``{The X-shooter pipeline},'' {\em Astronomische
  Nachrichten}~{\bf 332},  227 (Mar. 2011).

\end{thebibliography}
\bibliographystyle{spiebib} % makes bibtex use spiebib.bst
\end{document}